\documentclass[twoside,twocolumn,9pt]{article}
\usepackage{extsizes}
\usepackage[super,sort&compress,comma]{natbib} 
\usepackage[version=3]{mhchem}
\usepackage[left=1.5cm, right=1.5cm, top=1.785cm, bottom=2.0cm]{geometry}
\usepackage{balance}
\usepackage{mathptmx}
\usepackage{sectsty}
\usepackage{graphicx} 
\usepackage{lastpage}
\usepackage[format=plain,justification=justified,singlelinecheck=false,font={stretch=1.125,small,sf},labelfont=bf,labelsep=space]{caption}
\usepackage{float}
\usepackage{fancyhdr}
\usepackage{fnpos}
\usepackage[english]{babel}
\addto{\captionsenglish}{%
  
}
\usepackage{array}
\usepackage{droidsans}
\usepackage{charter}
\usepackage[T1]{fontenc}
\usepackage[usenames,dvipsnames]{xcolor}
\usepackage{setspace}
\usepackage[compact]{titlesec}
\usepackage{hyperref}

\usepackage{multirow}
\usepackage{epstopdf}%This line makes .eps figures into .pdf - please comment out if not required.

\usepackage{color, soul}

\usepackage{tabu}
\newcommand{\Tstrut}{\rule[.2\baselineskip]{0pt}{\baselineskip}}
\newcommand{\Bstrut}{\rule[-.5\baselineskip]{0pt}{\baselineskip}}

\definecolor{cream}{RGB}{222,217,201}

\begin{document}

\pagestyle{fancy}
\thispagestyle{plain}
\fancypagestyle{plain}{
%%%HEADER%%%
\renewcommand{\headrulewidth}{0pt}
}
%%%END OF HEADER%%%

%%%PAGE SETUP - Please do not change any commands within this section%%%
\makeFNbottom
\makeatletter
\renewcommand\LARGE{\@setfontsize\LARGE{15pt}{17}}
\renewcommand\Large{\@setfontsize\Large{12pt}{14}}
\renewcommand\large{\@setfontsize\large{10pt}{12}}
\renewcommand\footnotesize{\@setfontsize\footnotesize{7pt}{10}}
\makeatother

\renewcommand{\thefootnote}{\fnsymbol{footnote}}
\renewcommand\footnoterule{\vspace*{1pt}% 
\color{cream}\hrule width 3.5in height 0.4pt \color{black}\vspace*{5pt}} 
\setcounter{secnumdepth}{5}

\makeatletter 
\renewcommand\@biblabel[1]{#1}            
\renewcommand\@makefntext[1]% 
{\noindent\makebox[0pt][r]{\@thefnmark\,}#1}
\makeatother 
\renewcommand{\figurename}{\small{Fig.}~}
\sectionfont{\sffamily\Large}
\subsectionfont{\normalsize}
\subsubsectionfont{\bf}
\setstretch{1.125} %In particular, please do not alter this line.
\setlength{\skip\footins}{0.8cm}
\setlength{\footnotesep}{0.25cm}
\setlength{\jot}{10pt}
\titlespacing*{\section}{0pt}{4pt}{4pt}
\titlespacing*{\subsection}{0pt}{15pt}{1pt}
%%%END OF PAGE SETUP%%%

%%%FOOTER%%%
\fancyfoot{}
%\fancyfoot[LO,RE]{\vspace{-7.1pt}\includegraphics[height=9pt]{head_foot/LF}}
%\fancyfoot[CO]{\vspace{-7.1pt}\hspace{11.9cm}\includegraphics{head_foot/RF}}
%\fancyfoot[CE]{\vspace{-7.2pt}\hspace{-13.2cm}\includegraphics{head_foot/RF}}
\fancyfoot[RO]{\footnotesize{\sffamily{1--\pageref{LastPage} ~\textbar  \hspace{2pt}\thepage}}}
\fancyfoot[LE]{\footnotesize{\sffamily{\thepage~\textbar\hspace{4.65cm} 1--\pageref{LastPage}}}}
\fancyhead{}
\renewcommand{\headrulewidth}{0pt} 
\renewcommand{\footrulewidth}{0pt}
\setlength{\arrayrulewidth}{0.4pt}
\setlength{\columnsep}{6.5mm}
\setlength\bibsep{1pt}
%%%END OF FOOTER%%%

%%%FIGURE SETUP - please do not change any commands within this section%%%
\makeatletter 
\newlength{\figrulesep} 
\setlength{\figrulesep}{0.5\textfloatsep} 

\newcommand{\topfigrule}{\vspace*{-1pt}% 
\noindent{\color{cream}\rule[-\figrulesep]{\columnwidth}{1.5pt}} }

\newcommand{\botfigrule}{\vspace*{-2pt}% 
\noindent{\color{cream}\rule[\figrulesep]{\columnwidth}{1.5pt}} }

\newcommand{\dblfigrule}{\vspace*{-1pt}% 
\noindent{\color{cream}\rule[-\figrulesep]{\textwidth}{1.5pt}} }

\makeatother
%%%END OF FIGURE SETUP%%%

%%%TITLE, AUTHORS AND ABSTRACT%%%
\twocolumn[
  \begin{@twocolumnfalse}
{\hfill\raisebox{0pt}[0pt][0pt]\\}\par
\vspace{1em}
\sffamily
\begin{tabular}{m{4.5cm} p{13.5cm} }

 & \noindent\LARGE{X-ray induced desorption and photochemistry in CO ice} \\%Article title goes here instead of the text "This is the title"
\vspace{0.3cm} & \vspace{0.3cm} \\

 & \noindent\large{R. Dupuy,$^{\ast}$\textit{$^{a}$} M. Bertin,\textit{$^{a}$} G. F\'eraud\textit{$^{a}$}, C. Romanzin,\textit{$^{b}$} T. Putaud,\textit{$^{a}$} L. Philippe,\textit{$^{a}$} X. Michaut,\textit{$^{a}$}, P. Jeseck,\textit{$^{a}$} R. Cimino,\textit{$^{c}$} V. Baglin,\textit{$^{d}$} and J.-H. Fillion\textit{$^{a}$}} \\%Author names go here instead of "Full name", etc.

 & \noindent\normalsize{We report an investigation of X-ray induced desorption of neutrals, cations and anions from CO ice. The desorption of neutral CO, by far the most abundant, is quantified and discussed within the context of its application to astrochemistry. The desorption of many different cations, including large cations up to the mass limit of the spectrometer, are observed. In contrast, the only desorbing anions detected are O$^-$ and C$^-$. The desorption mechanisms of all these species are discussed with the aid of their photodesorption spectrum. The evolution of the X-ray absorption spectrum shows significant chemical modifications of the ice upon irradiation, which along with the desorption of large cations gives a new insight into X-ray induced photochemistry in CO ice.} \\%The abstrast goes here instead of the text "The abstract should be..."

\end{tabular}

 \end{@twocolumnfalse} \vspace{0.6cm}

  ]
%%%END OF TITLE, AUTHORS AND ABSTRACT%%%

%%%FONT SETUP - please do not change any commands within this section
\renewcommand*\rmdefault{bch}\normalfont\upshape
\rmfamily
\section*{}
\vspace{-1cm}

%%%FOOTNOTES%%%

\footnotetext{\textit{$^{a}$~Sorbonne Universit\'e, Observatoire de Paris, Universit\'e PSL, CNRS, LERMA, F-75005, Paris, France; E-mail: dupuy@fhi.mpg.de}}
\footnotetext{\textit{$^{b}$~Institut de Chimie Physique, UMR 8000, CNRS, Universit\'e Paris-Saclay, 91405, Orsay, France}}
\footnotetext{\textit{$^{c}$~Laboratori Nazionali di Frascati (LNF)-INFN I-00044 Frascati}}
\footnotetext{\textit{$^{d}$~CERN, CH-1211 Geneva 23, Switzerland}}

%Please use \dag to cite the ESI in the main text of the article.
%If you article does not have ESI please remove the the \dag symbol from the title and the footnotetext below.
%\footnotetext{\dag~Electronic Supplementary Information (ESI) available: [details of any supplementary information available should be included here]. See DOI: 10.1039/cXCP00000x/}
%additional addresses can be cited as above using the lower-case letters, c, d, e... If all authors are from the same address, no letter is required

%\footnotetext{\ddag~Additional footnotes to the title and authors can be included \textit{e.g.}\ `Present address:' or `These authors contributed equally to this work' as above using the symbols: \ddag, \textsection, and \P. Please place the appropriate symbol next to the author's name and include a \texttt{\textbackslash footnotetext} entry in the the correct place in the list.}

%%%END OF FOOTNOTES%%%

%%%MAIN TEXT%%%%

\section{Introduction}

Molecular ices composed of small molecules such as H$_2$O, CO or CH$_4$ are found at the surface of many cold astrophysical bodies, including comets\cite{mumma2011}, icy moons and planets of the outer solar system\cite{bennett2013}, but also micrometric dust grains of the interstellar medium\cite{boogert2015}. CO ice, for example, is abundantly found on Triton and Pluto\cite{bennett2013}, and in most comets\cite{mumma2011}, including recent detections of very high amounts of CO in the coma of an Oort cloud comet\cite{biver2018} and of the first detected interstellar comet, 2I/Borisov\cite{cordiner2020}. In cold regions of the interstellar medium, when the temperature drops below the freeze-out point of CO, all the gas phase CO molecules condense very rapidly at the surface of the dust grains, forming an outer icy layer composed mostly of this molecule\cite{boogert2015}.

These molecular ices are affected by many sources of irradiations, such as UV and X-ray photons from nearby (directional) and distant (background) stars, cosmic rays, and the secondary electrons and UV photons generated by these primary sources. Irradiation affects the structure and chemical composition of the ice mantle \cite{dartois2015a,rothard2017,oberg2016}, but also regulates the exchanges between ice and gas-phase, through conversion of the deposited energy (often in the form of electronic excitations) to desorption of molecules \cite{bennett2013}. Non-thermal desorption processes are crucial to link observed gas-phase molecule abundances to the inferred interplay of gas-phase and solid-phase chemistry that led to their formation\cite{huang2016}. A quantitative description of these processes along with a thorough understanding of the interaction between molecular ices and different types of irradiation are thus important. 

This need for data has led to a renewal of interest for non-thermal desorption, and in particular for VUV photodesorption, with many quantitative studies on the VUV photodesorption of CO\cite{oberg2009b,munozcaro2010,fayolle2011,bertin2012,chen2013a,paardekooper2016,diaz2019}
. While VUV photons are ubiquitous in the interstellar medium, they are not necessarily the dominant source of irradiation in every region. Cosmic-ray induced desorption from CO has also been studied from early on\cite{chrisey1986,chrisey1990} and more recently with heavy ions\cite{seperueloduarte2010}. In several recent papers, we have explored the astrophysical implications\cite{dupuy2018c} and the details\cite{dupuy2020} of the X-ray photodesorption process from water ice, as well as X-ray photodesorption from methanol in pure\cite{basalgete2021a} and mixed ices\cite{basalgete2021}. Other authors also studied X-ray photodesorption from mixed ices\cite{jimenez-escobar2018,ciaravella2020}. Through the study of water ice \cite{dupuy2018c}, we found that X-ray photodesorption is a potentially important process in some regions of the interstellar medium, comforting the need for quantitative studies on other relevant molecular ices. Each of the many desorbing species (neutrals, cations and anions) from water ice exhibited different behaviours, leading to a variety of possible desorption mechanisms and giving insights into the details of energy relaxation of X-ray excited water ice\cite{dupuy2020}. A similarly complete survey of desorbing species from CO ice is therefore also of interest. Early studies of X-ray photodesorption from CO ice\cite{rosenberg1985,scheuerer1990} were exploratory and only looked at cation desorption. Several X-ray desorption studies have been made for CO chemisorbed on various surfaces \cite{treichler1991,jugnet1984,martensson1990,weimar2000}, but such systems are very different from the condensed CO ice. 

In this paper, we study and quantify desorption of all species from CO ice, including neutrals and anions. We discuss the details of the photodesorption mechanisms for the different species, as well as what ion desorption and X-ray absorption spectroscopy reveal of CO X-ray photochemistry. Desorption of neutrals is by far the most abundant, and we derive astrophysically relevant yields for neutral CO X-ray photodesorption in the same way described in a previous paper\cite{dupuy2018c}.

\section{Methods}

\subsection{Experimental set-up}

Experiments were performed in the SPICES 2 set-up. Some aspects of the experiments have already been detailed elsewhere\cite{dupuy2018c,dupuy2020}. Briefly, the set-up is an ultra-high vacuum chamber equipped with a closed-cycle helium cryostat, reaching a base temperature of 15 K at the sample holder and a base pressure of $\sim 1 \times 10^{-10}$ mbar at 15 K. The substrate used was a technical copper surface (polycrystalline OFHC copper), electrically insulated from the holder by a kapton foil. CO (Airliquide) or isotopically labelled $^{13}$CO (Eurisotop, $>$99\% $^{13}$C purity) was injected through a dosing tube to grow a $\sim$ 100 monolayers (ML) thick ice on the substrate at 15 K at a rate of $\sim$0.2 ML s$^{-1}$. The thickness ensures a negligible substrate influence on desorption.

The set-up was installed on the SEXTANTS beamline of the SOLEIL synchrotron. During irradiation, the photon energy was scanned typically between 525 and 600 eV in steps of 0.2 eV. To better resolve fine structures near the edge, some scans used a step size of 0.05 eV instead. The monochromatized beam had a resolution of 80 meV and a flux of 7 $\times$ 10$^{11}$ photon s$^{-1}$ for most of the experiments. A higher flux was used for some specific experiments that are described in the text. The flux is constant over the whole photon energy range except around 535 eV where a dip is present. The spot at the surface was approximately 0.1 cm$^2$ large. The beam was incident at 47$^{\circ}$ relative to the surface normal, and the polarization was set to horizontal so that at the surface the light had a half out-of-plane and half in-plane components.

Absorption spectroscopy of the ice is performed by recording the current of the electrically insulated substrate that appears upon irradiation from the ejection of electrons from the surface (total electron yield, TEY), and scanning the photon energy. The absolute energy scale was set in a way described in a previous paper\cite{dupuy2020}. The position of the main resonance peak of solid CO is known to be the same as in the gas phase\cite{jugnet1984} and thus served as a cross-check of the validity of our energy scale. 

Neutral species desorption was detected using a quadrupole mass spectrometer (Pfeiffer Vacuum). Positive and negative ion desorption was detected with another quadrupole mass spectrometer (EQS Hiden Analytical). This QMS is equipped with a 45$^{\circ}$ deflector kinetic energy analyser\cite{hamers1998} with a fairly large resolution of $\sim$1.5 eV. All the spectra that we present are taken at a given kinetic energy (KE-differentiated spectra), the center of the KE distribution unless stated otherwise.

\subsection{Ice charging and ageing}

CO ice is subject to ageing (chemical and/or structural modifications of the ice) and charging (accumulation of charges in the ice - presumably positive, considering the number of electrons escaping the ice is greater than the number of positive ions). Ageing can affect the desorption yield of all species observed, with possible variations within a single scan. For ions, charging also has an effect because of the kinetic energy filter of the QMS. Charging will cause a shift of the kinetic energy distributions (KEDs), while the kinetic energy filter setting remains the same in the course of a scan (it is set by measuring the kinetic energy distribution of the ion(s) of interest before the scan and fixed at the maximum of the distribution). In the duration of a long scan (typically 520-600 eV) the kinetic energy distributions can shift enough to have a potentially visible effect on the spectrum. Both charging and ageing are responsible for some discrepancies observed between the absorption and the photodesorption spectra, e.g. in the form of a slowly decreasing desorption signal. These issues cannot be avoided completely, as a high enough flux is required to obtain a good desorption signal. KE distributions at different fixed photon energies were measured and integrated to check that the KE-differentiated spectra we present are not distorted compared to the "true" KE-integrated spectra. 

\subsection{Calibration of the photodesorption yields}

The derivation of absolute photodesorption yields for neutral species follows the method detailed in a previous publication\cite{dupuy2018c} on water ice. The method consists in calibrating the QMS current against the absolute flux of desorbing molecules in a temperature-programmed desorption (TPD) experiment where the amount of thermally desorbed molecules is known. 

To obtain relative ion desorption yields, we estimated the relative detection efficiency of the QMS for the different ions. We assume the relative detection efficiency only depends on the mass of the ion. Estimations made with different gases with known cracking patterns and compared using a calibrated pressure gauge allowed to derive an apparatus function for the QMS, which roughly follows a (m/z)$^{-0.5}$ power law. After correction by the apparatus function, the relative ion yields can be converted to absolute values by taking desorption of the C$^+$ ion from neat CO ice as 2 $\times$ 10$^{-5}$ desorbed ion/incident photon, which is the value obtained by Rosenberg et al.\cite{rosenberg1985} using their calibrated apparatus. Based on this measurement, the estimated absolute detection efficiency of our own QMS for this ion would be $\sim$2\%.

The uncertainty of neutral species calibration was estimated\cite{dupuy2018c} at $\pm$40\%. The absolute values for cations (and anions) rely on the validity of the measurements of Rosenberg et al. The relative comparison of cations, however, is valid within the $\pm$30\% error of the apparatus function calibration. The inter-comparison of cations and anions, on the other hand, relies on the assumption that the detection efficiency still only depends on the mass, which given that other parameters of the QMS change may not be true. The calibration of anions should therefore rather be considered as an order of magnitude estimate. 

\section{Results}

\subsection{Solid CO absorption spectroscopy}

\subsubsection{Spectral attributions.}

\begin{figure*}
	\centering
    \includegraphics[trim={0cm 0cm 0cm 0cm},clip,width=0.7\linewidth]{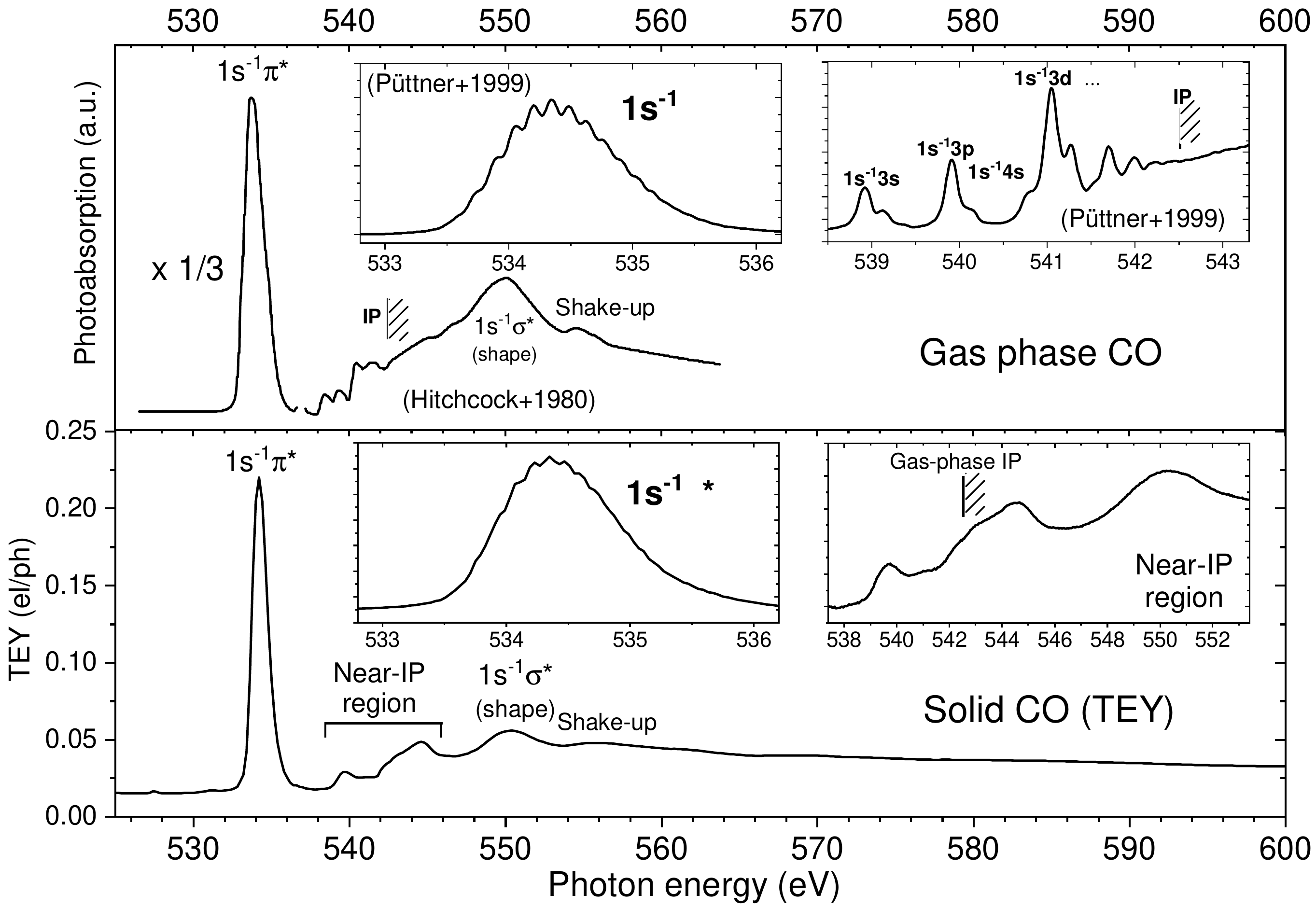}
    \caption{Photoabsorption spectrum of gas phase CO (upper panel) and neat solid CO (lower panel). The insets of both panels show spectra at higher resolution showing specific regions in more details. The top panel is the gas phase spectrum of CO. The moderate resolution spectrum of the main panel is adapted from Hitchcock et al.\cite{hitchcock1980}, while the insets display high resolution photoionisation data of specific regions, adapted from P\"uttner et al.\cite{puttner1999}. Some attributions are indicated on the figure. The lower panel is the solid phase spectrum of CO, from our TEY measurements. The main panel shows the full spectrum from 525 to 600 eV while the insets correspond to spectra of specific regions with a finer scan step. The spectral resolution of the light is $\sim$80 meV for our spectra.} 
    \label{CO_abs_spectrum}
\end{figure*} 

The absorption spectrum of neat solid CO near the O 1s edge, under our irradiation conditions, is displayed in fig. \ref{CO_abs_spectrum} (bottom panel). Spectra of specific regions taken with finer step size are shown in the insets as well. In order to interpret the absorption spectrum of solid CO, it is compared to the absorption spectrum of gas phase CO in the same region. A moderate resolution EELS spectrum of gas phase CO between 530 and 560 eV, taken from Hitchcock et al. \cite{hitchcock1980} is shown in the upper panel of fig. \ref{CO_abs_spectrum}, along with high resolution photoionisation spectra in specific regions taken from P\"uttner et al. \cite{puttner1999} (insets). The gas phase and solid phase spectra are very similar, with a few notable exceptions. Attribution of the gas phase near edge spectrum is relatively straightforward since the peaks consist in well-spaced transitions of a 1s electron to empty orbitals of CO. The transitions in the gas phase are vibrationally resolved, indicating that the core-hole states are bound. The first and strongest feature is the 1s$^{-1}\pi^*$ resonance (1$\sigma$-2$\pi$ transition), centered at 534.4 eV. The left inset of the upper panel of fig. \ref{CO_abs_spectrum} is a zoom onto the 1s$^{-1}\pi^*$ gas phase resonance with resolved vibrational peaks. An almost identical peak, with less well-resolved vibrational peaks but a similar width, is observed for solid CO (left inset of the bottom panel). Jugnet et al. \cite{jugnet1984} have imaged the 1s$^{-1}\pi^*$ resonance for chemisorbed, gas phase and solid CO, and already pointed out an exact match of the position and width of this peak for solid and gas phase CO. The core electronic state is thus remarkably unaffected by the condensed environment. This stands in contrast with what is observed in the case of water ice\cite{nilsson2010,dupuy2020}, highlighting the differences between hydrogen-bond solids and weakly interacting Van der Waals solids.

In the right inset of the upper panel in fig. \ref{CO_abs_spectrum}, a series of peaks are attributed to transitions to successive Rydberg states, with again resolved vibrational peaks. The ionization potential (IP) of gas phase CO is 542.54 eV and we see clearly the convergence of the Rydberg states to this threshold. This region is the most modified in the condensed phase, where we see (right inset of the bottom panel, fig. \ref{CO_abs_spectrum}) a series of more or less well resolved broad features. Vibrational progressions are no longer visible. There is a clear feature at 539.9 eV, another at 544.7 eV, and at least two shoulders around 541.4 eV and 543.4 eV, with possibly more, broader peaks contributing here. As expected, Rydberg states are considerably modified in the condensed phase, because of their large spatial extent overlapping with neighbouring molecules, and thus very broad peaks are observed. 

Contrary to the gas phase, the ionization threshold is not visible through the convergence of Rydberg states. No photoemission experiments on solid CO exist in this region to our knowledge. The solid phase IP is expected to be lower than the gas phase IP, as the ionic state is stabilized in the condensed phase due to polarisation screening. For example, the first IP of CO is red-shifted of $\sim$1.5 eV in the VUV region \cite{jacobi1982}. Unattributed features are also present in the gas phase spectrum of Hitchcock et al. (fig. \ref{CO_abs_spectrum} upper panel) in this region (542-548 eV), and the broad peaks observed in condensed phase could simply be inherited from these gas phase features that may correspond to multielectron excitation states (shake-up, shake-off...). They could also be specific to the condensed phase, being for example multiple scattering resonances, as suggested by Scheuerer et al.\cite{scheuerer1990}.

Still above in energy in the gas phase spectrum, there is a strong shape resonance at 550 eV which is inherited as is in the condensed phase. An additional weak feature at 556 eV, and possibly even weaker shoulders higher in energy, can also be observed. A similar weak feature is observed in the gas phase. This feature can most probably be attributed to a second, weaker shape resonance, and is discussed later in the text. 

\subsubsection{Evolution of the spectrum under irradiation.}

\begin{figure}
    \includegraphics[trim={0cm 0cm 0cm 0cm},clip,width=\linewidth]{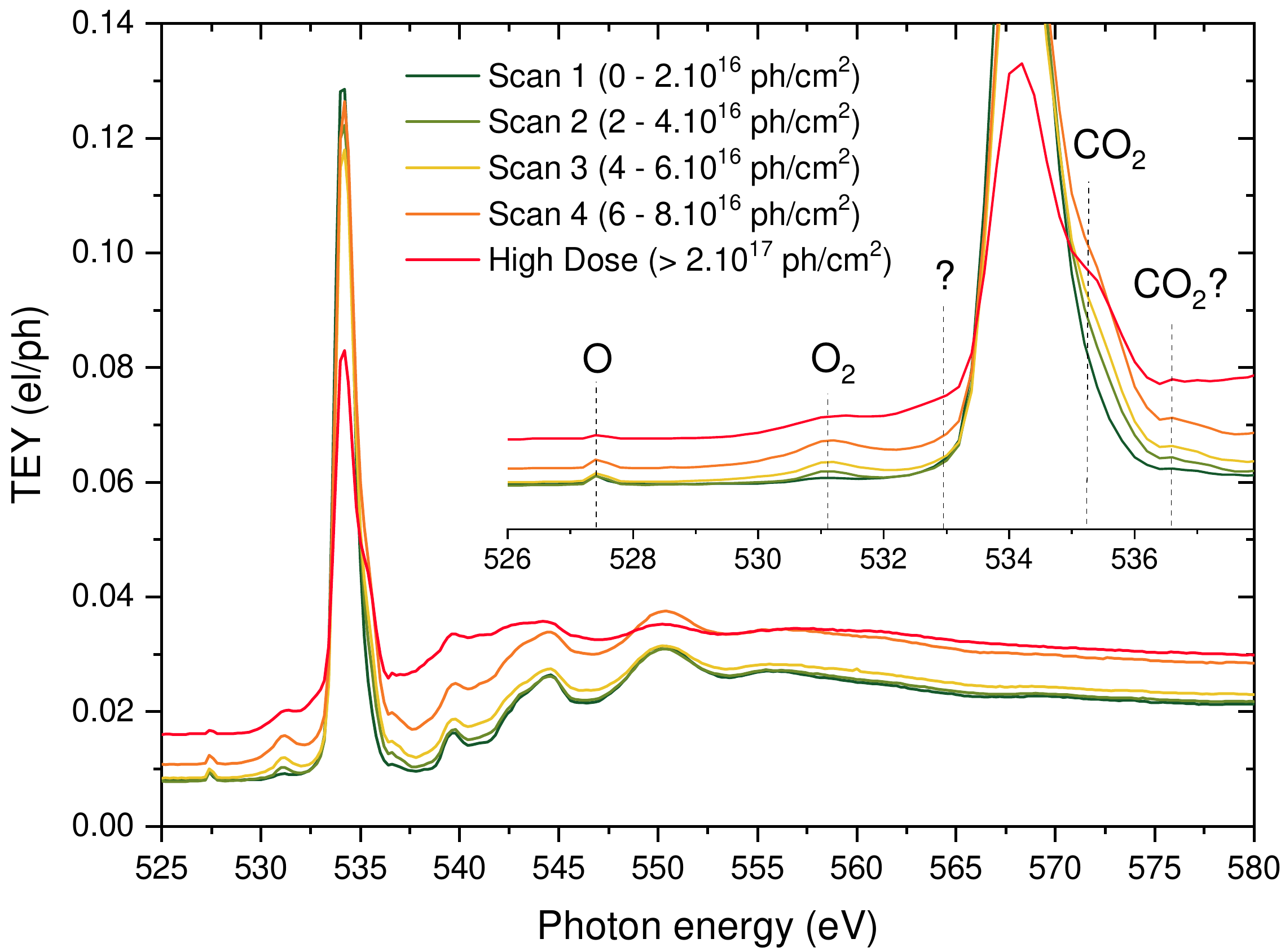}
    \caption{Evolution of the total electron yield (TEY) of solid CO with photon dose. Total electron yield for four consecutive scans on an initially fresh ice (scans labeled 1 to 4), photon dose $\sim$ $2\times 10^{16}$ photons cm$^{-2}$ per scan, and total electron yield of an ice having received a much higher photon dose ($>$ $2\times 10^{17}$ photons cm$^{-2}$; scan labeled "high dose"). The inset is a zoom in the region where photoproducts peaks can be seen. These peaks are attributed on the figure.} 
    \label{CO_TEYs}
\end{figure} 

Under the high photon flux conditions necessary to observe photodesorption processes, the ice is modified by photochemistry, which is reflected in the absorption spectrum. Even the first scan made on a neat CO ice already probes a slightly modified ice, which is visible for example in fig. \ref{CO_abs_spectrum} (bottom panel) by the presence of small peaks before the 1s$^{-1}\pi^*$ peak, that can be attributed to products of irradiation. In our previous studies on water ice\cite{dupuy2018c,dupuy2020}, the irradiation-induced chemistry reached a steady state already at the photon dose required for a single scan, meaning that the absorption spectrum of water ice no longer evolved upon further irradiation. Pure water ice showed a high resilience to irradiation, with most of the chemistry leading back to H$_2$O formation. 

CO ice does not reach such a chemical steady state during the irradiation. In fig. \ref{CO_TEYs}, the absorption spectra of a CO ice upon four successive scans ($\sim$ $2\times 10^{16}$ photons cm$^{-2}$ per scan) are shown, as well as the spectrum of a CO ice having received a very high photon dose ($>$ $2\times 10^{17}$ photons cm$^{-2}$). The overall shape of the TEY changes with photon dose: (i) the 1s$^{-1}\pi^*$ peak decreases, (ii) the baseline increases and (iii) the various features in the 539-550 eV region become progressively blurred, and are almost lost for the high dose ice. In addition, peaks in the pre-edge region corresponding to the photoproducts increase in intensity. Two of these peaks, at 527.4 and 531.2 eV, can respectively be attributed to atomic O and molecular O$_2$\cite{laffon2006,dupuy2020} (inset of fig. \ref{CO_TEYs}). The highest photoproduct peak is at 535.2 eV, in the right shoulder of the 1s$^{-1}\pi^*$ peak of CO, and can be attributed to CO$_2$ (based on the gas phase position of the 1s$^{-1}\pi^*$ resonance of CO$_2$ \cite{wight1974}). This resonance features a well-identified high-energy shoulder in the gas phase\cite{piancastelli1997} to which we can presumably attribute the feature growing at 536.6 eV in our spectra. At the highest irradiation dose, other broad features from photoproducts that are not precisely identified seem to contribute as well. The progressive blurring of the features in the near-IP region is due to the fact that all these photoproducts will contribute to absorption in this region, with all the different broad features and ionization continua blending together. This prevents any specific attribution in this region.

%We can rule out the possibility of a second peak corresponding to O, O$_2$ or CO$_2$, as none of the other resonances of these species are expected in this region. But the variety of molecules combining carbon and oxygen that can be formed is very large, as we will see later. It is also possible that these peaks corresponds to the absorption of a class of several similar molecules. According to the other evidence we have from desorption and other photochemistry experiments in the literature (to be discussed later), the most likely candidates are C$_2$O and C$_3$O$_2$, and possibly also C$_3$O. None of these molecules, which are not easily isolated in normal conditions, have available core excitation spectra, so that a definitive attribution is not possible. 

From the evolution of the TEYs, it is clear that the ice does not reach a steady state, at least not before a very high dose of photons. The modifications remain modest for the first four scans displayed in fig. \ref{CO_TEYs}, but photoproducts such as CO$_2$ and O$_2$ accumulate in the ice, as observed from the increase of their features in the absorption spectra. For ices having received a very high photon dose, the peaks of the photoproducts become blurred and it is not completely clear that their intensity has evolved much. The evidence that the ice still evolves then comes from the large changes in the near-IP region and the increase of the baseline. 

\subsection{Desorption of neutral species}

\begin{figure}
	\centering
    \includegraphics[trim={0cm 0cm 0cm 0cm},clip,width=\linewidth]{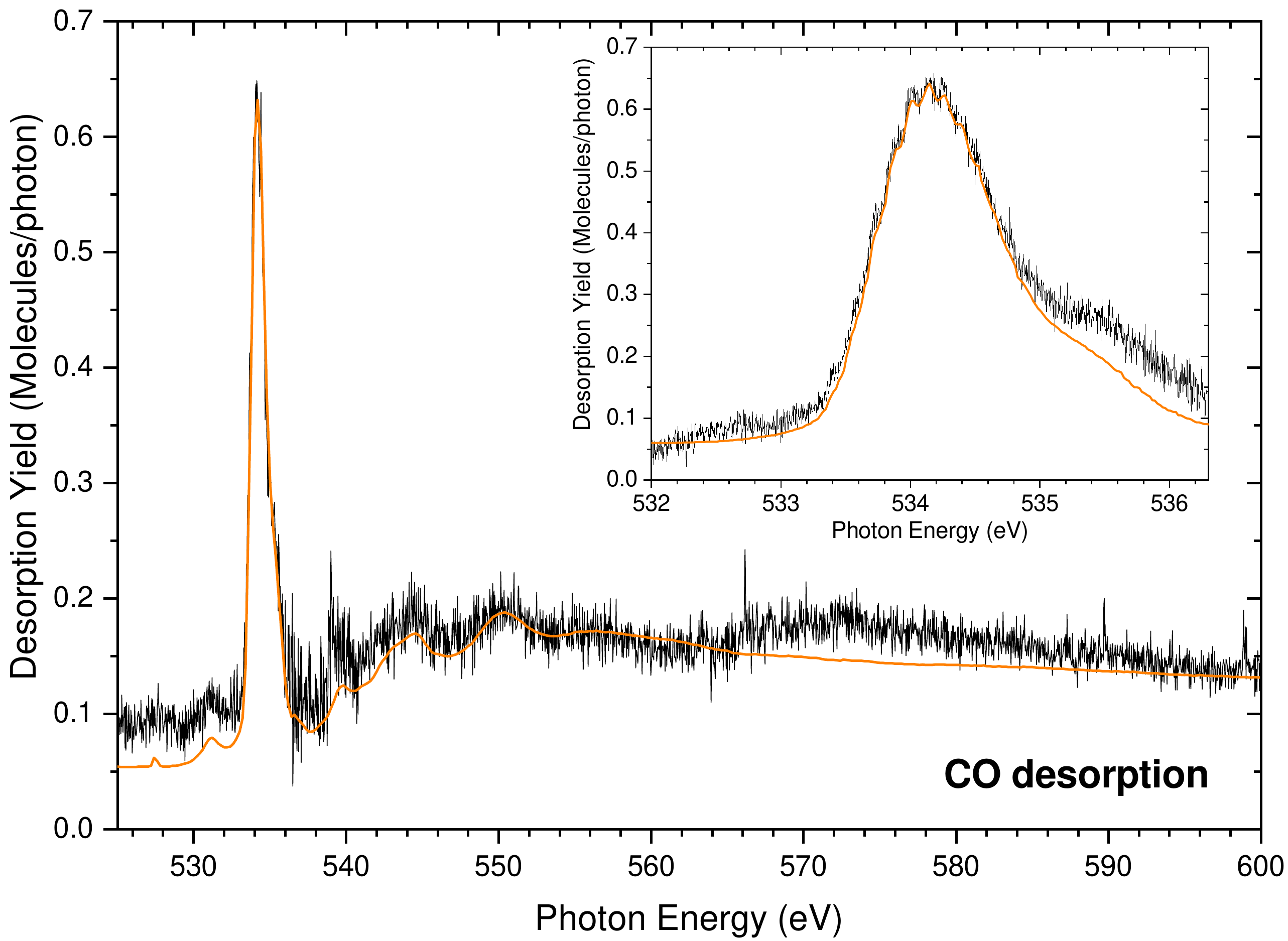}
    \caption{Photodesorption spectrum of neutral CO from solid CO near the O 1s edge. Added for comparison is the TEY (orange line, normalized so as to roughly match the photodesorption spectrum). This ice had already received a photon dose of $\sim$ $6-8 \times 10^{16}$ photons cm$^{-2}$. Inset: photodesorption spectrum around the 1s$^{-1}\pi^*$ resonance, from another ice. The vibrational structure is resolved. This ice had already been significantly irradiated ($>$ $1 \times 10^{17}$ photons cm$^{-2}$).} 
    \label{CO_psd_X}
\end{figure}

The photodesorption spectrum of the CO neutral molecule from solid CO in the O 1s region is shown in fig. \ref{CO_psd_X}. This spectrum was taken for an ice that had already received a medium photon dose (scan 4 of fig. \ref{CO_TEYs}, 6-8.10$^{16}$ photons cm$^{-2}$). The TEY for this ice shows nonetheless that it was not too heavily modified and remains mostly constituted of CO. Overall the spectrum, which is compared to the TEY (recorded simultaneously) on the figure, matches well with the absorption. The main features such as the 1s$^{-1}\pi^*$ resonance and the CO$_2$ shoulder, the O$_2$ peak, the near-IP region and the shape resonance are all recognizable and match with the absorption in relative intensities. Some observed discrepancies (behaviour of the baseline around 525 eV and 570 eV), are ascribed to background subtraction issues, which are common when looking at the desorption of the main ice molecule, as residual gas is introduced in the chamber upon ice dosing. Glitches with intensity bursts may also happen in the QMS (e.g. at 540 eV here). The inset shows a scan made with finer steps around the 1s$^{-1}\pi^*$ resonance, where we can see the vibrational features also on the desorption spectrum. Background issues definitely hinder the comparison between TEY and photodesorption spectrum here as well, and we do not assign physical meaning to the apparent difference between absorption and photodesorption at the CO$_2$ shoulder.  

\begin{figure}
	\centering
    \includegraphics[trim={0cm 0cm 0cm 0cm},clip,width=\linewidth]{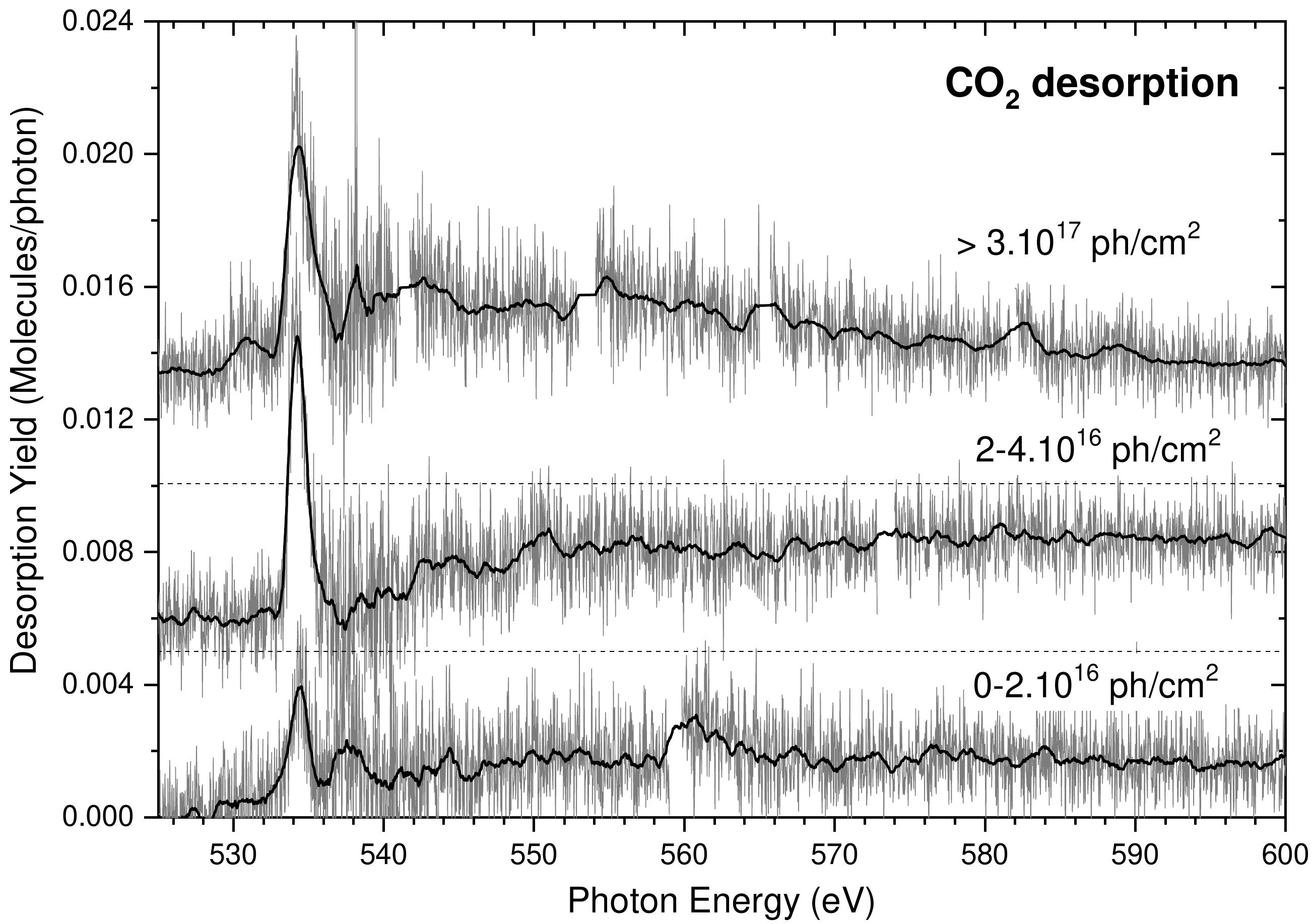}
    \caption{Photodesorption spectrum of neutral CO$_2$ from solid CO near the O 1s edge. Spectra for three different ices are presented (in grey is the raw data, the black line is a smooth). They have been vertically shifted for clarity; their baseline is indicated by the dashed lines. The bottom spectrum is for a fresh ice, the middle one was done immediately afterwards, and the upper one was done on another ice, that had already received a high photon dose ($>$ $3\times 10^{17}$ photons cm$^{-2}$). Cuts in the spectra (e.g. at 553 eV for the upper spectrum) are due to spurious intensity bursts of the QMS.} 
    \label{CO2_psd_X}
\end{figure}

The other neutral species that we could observe desorbing is CO$_2$. Fig. \ref{CO2_psd_X} shows three photodesorption spectra of CO$_2$: the first one was taken on a fresh ice, the second one immediately afterwards on the same ice, and the third one on another ice, that had been highly irradiated previously. Except the 1s$^{-1}\pi^*$ resonance and its shoulder, the features of the CO absorption spectrum are hardly recognizable because of the noise. The spectrum nonetheless seems to follow the absorption. We can see evolutions in the spectra with photon dose: from the first to the second scan, the desorption yield at the resonance more than doubles. The yields in the continuum, whether at 525 eV or 600 eV, increase as well (of a similar factor at 600 eV: more than doubled).  In the third scan, the yield at the resonance decreased, but so did the absorption (see the behaviour of highly irradiated ices in fig. \ref{CO_TEYs}). The yields at 525 eV and 600 eV, on the other hand, increased. Therefore on average over the whole energy range, CO$_2$ desorption (per absorbed photon) increases with time.

Using a much increased photon flux ($\sim 1.4 \times 10^{13}$ ph s$^{-1}$), we attempted to observe desorption of other neutral species. Search for desorption of O$_2$ and atomic O and C remained unsuccessful, despite the first two species being detected within the ice by absorption spectroscopy. However, the detection limit is relatively high on these mass channels (estimated around $< 1\times 10^{-3}$ molec ph$^{-1}$ at the 1s$^{-1}\pi^*$). The only other photodesorbed neutral species that was unambiguously detected is C$_2$O, with a desorption yield of the order of $1 \times 10^{-4}$ molec ph$^{-1}$ at 550 eV.

The absolute photodesorption yield of neutral CO in the O 1s edge region (0.64 mol.ph$^{-1}$ at the 1s$^{-1}\pi^*$, 0.15 molec ph$^{-1}$ at 600 eV) is high - more than an order of magnitude higher than H$_2$O desorption from water ice, for example\cite{dupuy2018c} -, but not much higher than the yield of CO in the VUV region\cite{fayolle2011} ($\sim$ 0.05 molec ph$^{-1}$ at 8.1 eV). This is partially explained by the much lower absorption cross-section of core-electron excitations compared with valence-electron ones. A more detailed comparison will be made in the Discussion.  

The absolute photodesorption yield of neutral CO$_2$ from solid CO is initially around  $1\times 10^{-3}$ molecule per incident photon at 600 eV and goes up to $4\times 10^{-3}$ after some irradiation. There is therefore almost two orders of magnitude between the yields. This CO$_2$ to CO desorption ratio is similar to the one found for electron-stimulated desorption from solid CO\cite{dupuy2020b}. On the other hand, CO$_2$ desorption is not observed during the VUV irradiation of solid CO, although small amounts are formed in the ice. 

\subsection{Desorption of ions}

\subsubsection{Mass spectrum of cations.}

\begin{figure}
	\centering
    \includegraphics[trim={0cm 1.5cm 0cm 1.5cm},clip,width=\linewidth]{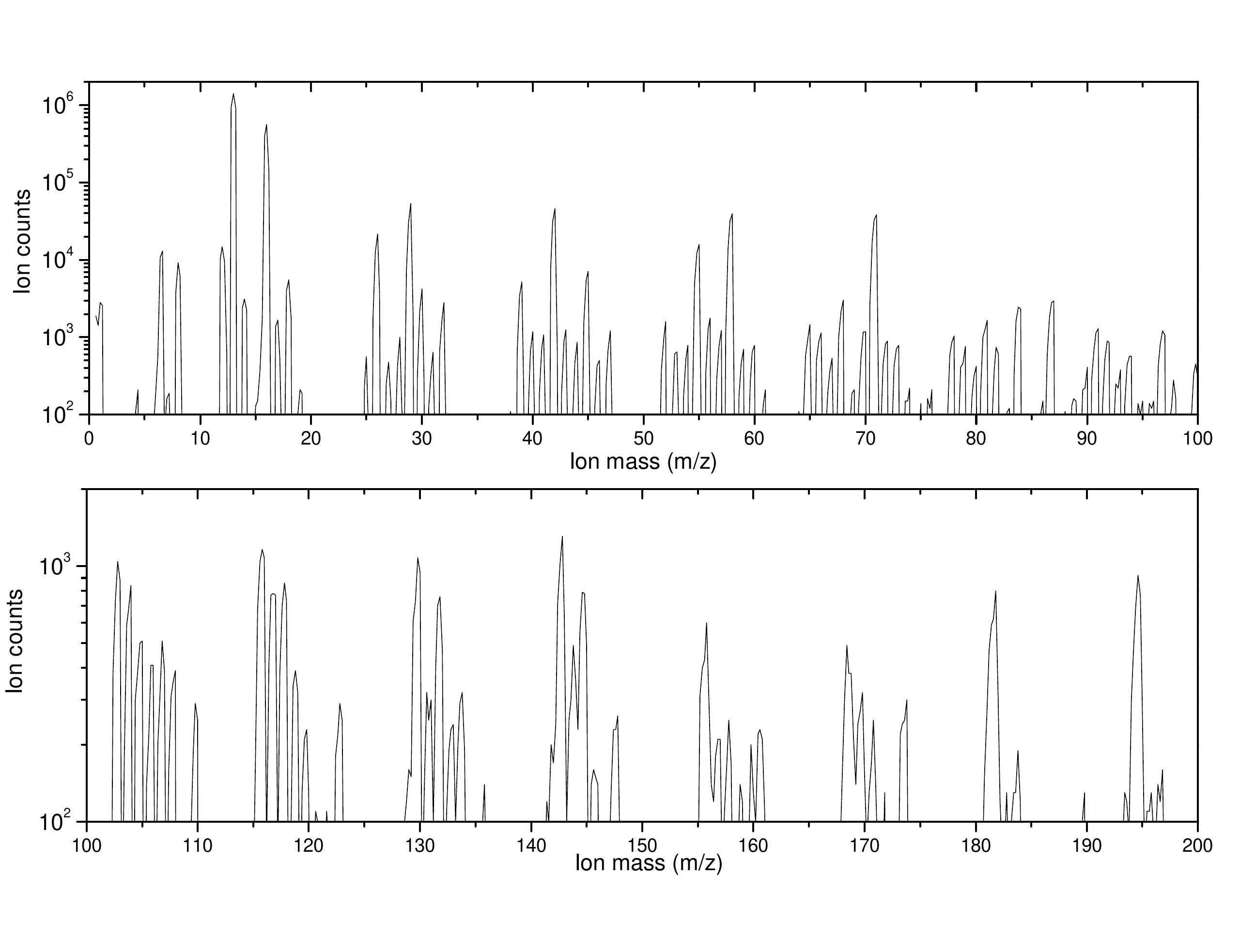}
    \caption{Mass spectrum of the cations desorbed from solid $^{13}$CO by irradiation at 550 eV. The spectrum was accumulated for a duration of $\sim$ 300 s with a photon flux of $2.3\times 10^{12}$ ph s$^{-1}$. The kinetic energy filter was set to 1.5 eV.} 
    \label{mass_spec_13CO}
\end{figure}

A very large number of cations are observed to desorb from X-ray irradiated CO ice, as shown in table \ref{mspec_CO_1-32} and \ref{mspec_CO_32+}. In contrast, only two anions, O$^-$ and C$^-$, were detected. The measured cation mass spectrum for $^{13}$CO ice (all mentions of C afterwards implicitly indicate $^{13}$C unless stated otherwise) irradiated at 550 eV is shown in fig. \ref{mass_spec_13CO}. Although C$^+$ and O$^+$ ions by far dominate the spectrum, other cations are observed up to our detection limit of m/z = 200. We will now discuss the attributions and intensities of the different measured peaks. For each peak we calculated the intensity relative to the $^{13}$C$^+$ (m/z = 13) peak corrected by the apparatus function of the QMS. 

\begin{table*}
\caption{Attribution and intensity of the desorbed cations from solid CO at 550 eV, between mass 1 and 32}
\def\arraystretch{1}
\label{mspec_CO_1-32}
\begin{tabu} to \textwidth {*3{X[c]} | *3{X[c]}}

Mass channel (a.m.u.) & Attribution$^*$ & Intensity rel. to $^{13}$C$^+$ & Mass channel (a.m.u.) & Attribution$^*$ & Intensity rel. to $^{13}$C$^+$ \Bstrut \\
\hline
\hline

1   & H$^+$               & 2     & 18 & $^{18}$O$^+$              & 0.46 \Tstrut\\ 
4.3 & C$^{3+}$            & 0.01  & 25 & $^{12}$CC$^+$             & 0.05 \\
6.5 & C$^{2+}$            & 0.67  & 26 & C$_2^+$                   & 2.14 \\
7   & N$^{2+}$            & 0.01  & 27 & HC$_2^+$                  & 0.05 \\
8   & O$^{2+}$            & 0.52  & 28 & $^{12}$CO$^+$             & 0.10 \\
12  & $^{12}$C$^+$        & 1.01  & 29 & CO$^+$                    & 5.57 \\
13  & C$^+$               & 100   & 30 & HCO$^+$                   & 0.45 \\
14  & N$^+$/CH$^+$        & 0.23  & 31 & H$_2$CO$^+$/C$^{18}$O$^+$ & 0.07 \\
16  & O$^+$               & 44.2  & 32 & O$_2^+$                   & 0.31 \\
17  & $^{17}$O$^+$/OH$^+$ & 0.13  &    &                           &      \Bstrut \\

\hline 

\end{tabu}
\footnotesize $^*$ All carbon atoms are $^{13}$C isotopes unless indicated otherwise. 
\end{table*}

Let us focus first on the part of the spectrum before mass 32, which concerns fragments and simple species. The peaks are exhaustively listed in table \ref{mspec_CO_1-32} along with their attribution and their intensity relative to C$^+$ ($^{13}$C$^+$ here). The most abundant fragments are C$^+$ and O$^+$. The CO$^+$ ion is much weaker, at 6\% of C$^+$. The only other peak with significant ($>$ 1\%) intensity is C$_2^+$. We also see an H$^+$ signal, indicating the presence of a few hydrogen pollutants. In the region of the multiply-charged atoms C$^{3+}$, C$^{2+}$ and O$^{2+}$ are observed, but also a faint trace of N$^{2+}$, indicating a very weak nitrogen pollution as well (most probably originating from previous use of ammonia). This is also visible on the weak N$^+$ (m/z =14) signal - although CH$^+$ could contribute to this mass as well, which makes it less reliable. There does not seem to be a CO$^{2+}$ signal contributing at m/z = 14.5. Because we used isotopic $^{13}$CO here, we can expect other isotopes of carbon and oxygen to be present at the percent level as well. This is why the signal at m/z = 12 is attributed to $^{12}$C$^+$, and the signals at m/z = 17 and 18 to $^{17}$O$^+$ and $^{18}$O$^+$ respectively. The latter two attributions are arguable because they could also correspond to pollutants such as OH$^+$ or H$_2$O$^+$. For this same reason, we can also expect isotopic variants of abundant ions to appear, such as the very weak signal at m/z = 25 which is attributed to $^{12}$C$^{13}$C$^+$, and the $^{12}$CO$^+$ signal. 

We can remark that it is not possible to cover all of the observed masses by simply making linear combinations of C and O atoms. Some peaks can be accounted for by the small presence of isotopic forms of C and O atoms and the possibility of doubly-ionized species, as explained above. For the other peaks, the conspicuous signal of H$^+$, along with the expected pollutants in the chamber (mainly H$_2$ and H$_2$O), suggest that hydrogen is involved and we see hydrogenated forms of C$_x$O$_y^+$ chains. Indeed, almost all of the peaks observed can be accounted for once the small presence of isotopes and singly hydrogenated HC$_x$O$_y^+$ ions are considered. In the region from mass 1 to 32, we can identify HC$_2^+$, HCO$^+$ and H$_2$CO$^+$, although C$^{18}$O$^+$ can also contribute to the latter signal (m/z = 31). The presence of hydrogenated pollutants remains very small compared to the total signal of non-hydrogenated ions originating from CO ionization, which is a testimony of the sensitivity of ion desorption rather than of an unusually polluted ice. 

\begin{table*}
\caption{Attribution and intensity of the the most intense desorbed cations from solid CO at 550 eV, between mass 33 and 200. Hydrogenated contaminants are excluded.}
\def\arraystretch{1.2}
\label{mspec_CO_32+}
\begin{tabu} to \textwidth {*3{X[c]}|*3{X[c]}}

Mass channel (a.m.u.) & Attribution$^*$ & Intensity rel. to $^{13}$C$^+$ & Mass channel (a.m.u.) & Attribution$^*$ & Intensity rel. to $^{13}$C$^+$  \Bstrut \\

\hline
\hline 

39  & C$_3^+$  & 0.63  & 42  & C$_2$O$^+$ & 5.76 \Tstrut\\
52  & C$_4^+$  & 0.22  & 55  & C$_3$O$^+$ & 2.24 \\
65  & C$_5^+$  & 0.22  & 68  & C$_4$O$^+$ & 0.47 \\
78  & C$_6^+$  & 0.17  & 81  & C$_5$O$^+$ & 0.28 \\
91  & C$_7^+$  & 0.23  & 94  & C$_6$O$^+$ & 0.10 \\
104 & C$_8^+$  & 0.16  & 107 & C$_7$O$^+$ & 0.10 \\
117 & C$_9^+$  & 0.16  & 120 & C$_8$O$^+$ & 0.05 \\
130 & C$_{10}^+$ & 0.23  & 133 & C$_9$O$^+$ & 0.05 \\
143 & C$_{11}^+$ & 0.29  & 146 & C$_{10}$O$^+$ & 0.04 \\
156 & C$_{12}^+$ & 0.14  & 159 & C$_{11}$O$^+$ & 0.03 \\
169 & C$_{13}^+$ & 0.12  &     &            &      \\
182 & C$_{14}^+$ & 0.20  &     &            &      \\
195 & C$_{15}^+$ & 0.24  &     &            &      \Bstrut \\
 
\hline

45  & CO$_2^+$      & 0.92  & 106  & C$_2$O$_5^+$ & 0.08 \Tstrut \\
71  & C$_3$O$_2^+$  & 6.12  & 119  & C$_3$O$_5^+$ & 0.08 \\
84  & C$_4$O$_2^+$  & 0.43  & 132  & C$_4$O$_5^+$ & 0.16 \\
97  & C$_5$O$_2^+$  & 0.23  & 158  & C$_6$O$_5^+$ & 0.06 \\
110 & C$_6$O$_2^+$  & 0.06  & 171  & C$_7$O$_5^+$ & 0.06 \\
    & 			    &       & 184  & C$_8$O$_5^+$ & 0.05 \\
    &   			&       & 197  & C$_9$O$_5^+$ & 0.04 \Bstrut \\
 
\hline 

58  & (CO)$_2^+$  & 5.78  & 100  & C$_4$O$_3^+$ & 0.08 \Tstrut\\
87  & (CO)$_3^+$  & 0.52  & 103  & C$_3$O$_4^+$ & 0.20 \\
116 & (CO)$_4^+$  & 0.24  & 148  & C$_4$O$_6^+$ & 0.06 \\
145 & (CO)$_5^+$  & 0.18  & 161  & C$_5$O$_6^+$ & 0.05 \Bstrut\\
 
\hline 

\end{tabu}
\footnotesize $^*$ All carbon atoms are $^{13}$C isotopes unless indicated otherwise. 

\end{table*}

The heavier cations are listed in table \ref{mspec_CO_32+}. The list is not exhaustive: as mentioned previously, isotopic and hydrogenated variants need to be considered to attribute all peaks. Pure CO chemistry remains dominant: above m/z=32, if we were to set an arbitrary limit and look only at peaks whose intensity are $>$ 0.2\% of the C$^+$ peak, only one peak (m/z = 56, 0.25\% of C$^+$) cannot be attributed to a species of the form C$_x$O$_y^+$. Only the C$_x$O$_y^+$ ions that are detected are shown in table \ref{mspec_CO_32+}. They are classified by family, depending on the number of oxygen atoms: we detect C$_x^+$ for x = 2 to 15, C$_x$O$^+$ for x = 2 to 11, C$_x$O$_2^+$ for x = 1 to 6, (CO)$_x^+$ for x = 2 to 5, and C$_x$O$_5^+$ for x = 2 to 9. Outside a clearly identifiable family we observe C$_4$O$_3^+$, C$_3$O$_4^+$, C$_4$O$_6^+$ and C$_5$O$_6^+$. It should be noted that we only have access to the chemical formula of the compound here, without any information on the way the atoms are arranged and if there are weakly bound ion clusters included (for example, the (CO)$_x^+$ are assumed to form a series of singly charged CO clusters, but that is not necessarily the case). The ions outside a series may correspond to particularly stable/abundant clusters. Although not shown in the table, the hydrogenated form of the carbon chains HC$_x^+$ were systematically found in the spectrum, with an intensity close to that of the corresponding C$_x^+$ peak. For the other hydrogenated variants the intensity is about an order of magnitude lower than the main peak. 

Four ions show significantly higher intensity than the rest: C$_2$O$^+$ (6\%), C$_3$O$^+$(2\%), (CO)$_2^+$ (6\%) and C$_3$O$_2^+$ (6\%). All other peaks are in the 0.05-1\% range. This is another argument for the C$_2$O or C$_3$O$_2$ neutral forms being the most likely candidates in the attribution of the unknown peaks in the absorption (see section 3.1.2). Surprisingly, the CO$_2^+$ ion is not very intense, despite being the most abundant neutral product after CO. The evolution of the intensity of the series of ions is interesting: while the C$_x$O$^+$, C$_x$O$_2^+$ and (CO)$_x^+$ show a clear decrease with increasing x, this is not the case for the C$_x^+$ and C$_x$O$_5^+$ series. The interpretation of these trends is unclear for now. 

The first study of X-ray induced ion desorption from CO ice by Rosenberg et al. \cite{rosenberg1985} only reported C$^+$, O$^+$, CO$^+$, C$_2$O$^+$ and (CO)$_2^+$ desorption, but a lot of the cations we detected here are also reported in a conference proceedings by Scheuerer et al.\cite{scheuerer1990}. Most of the ions were also seen in ion bombardment (1 keV Ar$^+$) of solid CO \cite{jonkman1981}. The relative intensities reported in these works all seem in reasonable agreement with ours.

\subsubsection{Spectral signatures.}

The spectra of some selected, abundant cations among all those observed were investigated. Here we will discuss the photodesorption spectra of C$^+$, O$^+$ and CO$^+$, as well as O$^-$. Spectra of C$_2^+$, C$_2$O$^+$, C$_3$O$^+$, C$_3$O$_2^+$ and (CO)$_2^+$ were taken (not shown) and show relatively little deviation from the absorption spectrum, within the limits of the signal/noise ratio which is low compared with the spectra presented below. The interpretation of all the spectra presented here needs to take into account the problems of charging and ageing mentioned in the Methods.  

\begin{figure}
	\centering
    \includegraphics[trim={0cm 0cm 0cm 0cm},clip,width=\linewidth]{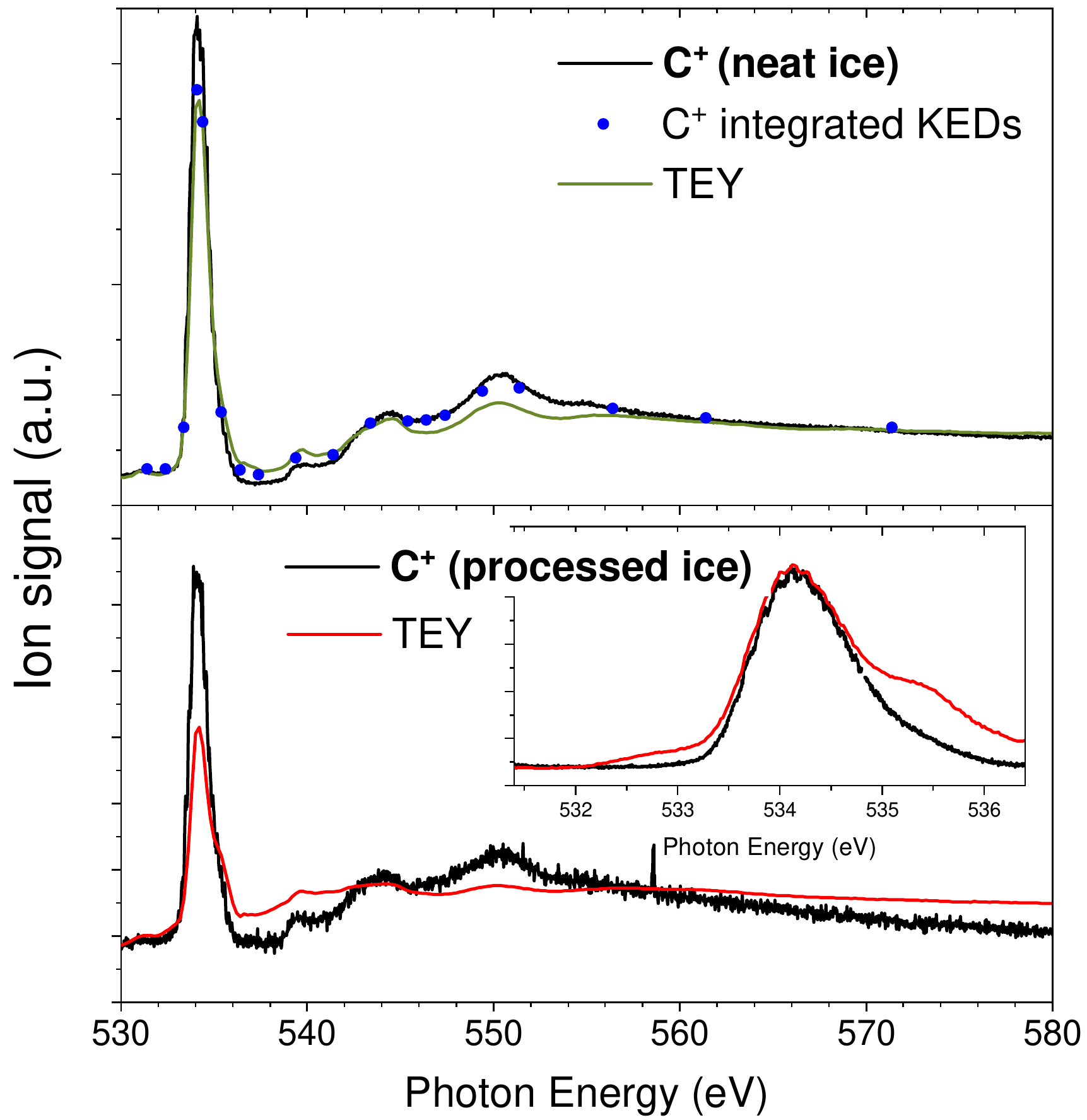}
    \caption{Photodesorption spectrum of C$^+$ for a neat ice (top panel) and a highly processed ice (bottom panel). Also shown on the top panel are the integrated KEDs for C$^+$ (blue dots) and the TEY (green line). The TEY is also shown in red in the bottom panel. The inset of the bottom panel is a zoom in the 1s$^{-1}\pi^*$ region, with the TEY scaled differently for comparison.} 
    \label{C+_psd}
\end{figure}

The photodesorption spectrum of C$^+$ from a neat ice (corresponding to the TEY scan 2 of fig. \ref{CO_TEYs}) is presented in the top panel of fig. \ref{C+_psd} along with the integrated KEDs taken at fixed photon energies (blue dots). The absorption is shown for comparison on the figure. Overall the features of CO absorption are all clearly seen in the photodesorption spectrum, and no significant deviation can be observed. The slope of the decrease of the continuum is higher on the photodesorption spectrum than on the TEY, which is  attributed to an ageing effect.The integrated KEDs confirm that the C$^+$ desorption spectrum follows the absorption. The bottom panel of fig. \ref{C+_psd} shows the photodesorption spectrum of C$^+$ from a highly irradiated ice. The inset is a spectrum with finer steps around the 1s$^{-1}\pi^*$ resonance. No significant deviations can be observed between photodesorption and absorption spectra. The C$^+$ photodesorption spectrum resembles in fact much more to the neat ice absorption and C$^+$ desorption spectra than to the highly irradiated ice absorption spectrum. All the CO features are clearly visible in the photodesorption spectrum, while they are blurred in the absorption spectrum, and the CO$_2$ contribution near the $\pi^*$ resonance is almost not visible. This suggests a much more efficient C$^+$ desorption upon core excitation of CO than upon core excitation of CO$_2$ or other species present in the ice.

\begin{figure}
	\centering
    \includegraphics[trim={0cm 0cm 0cm 0cm},clip,width=\linewidth]{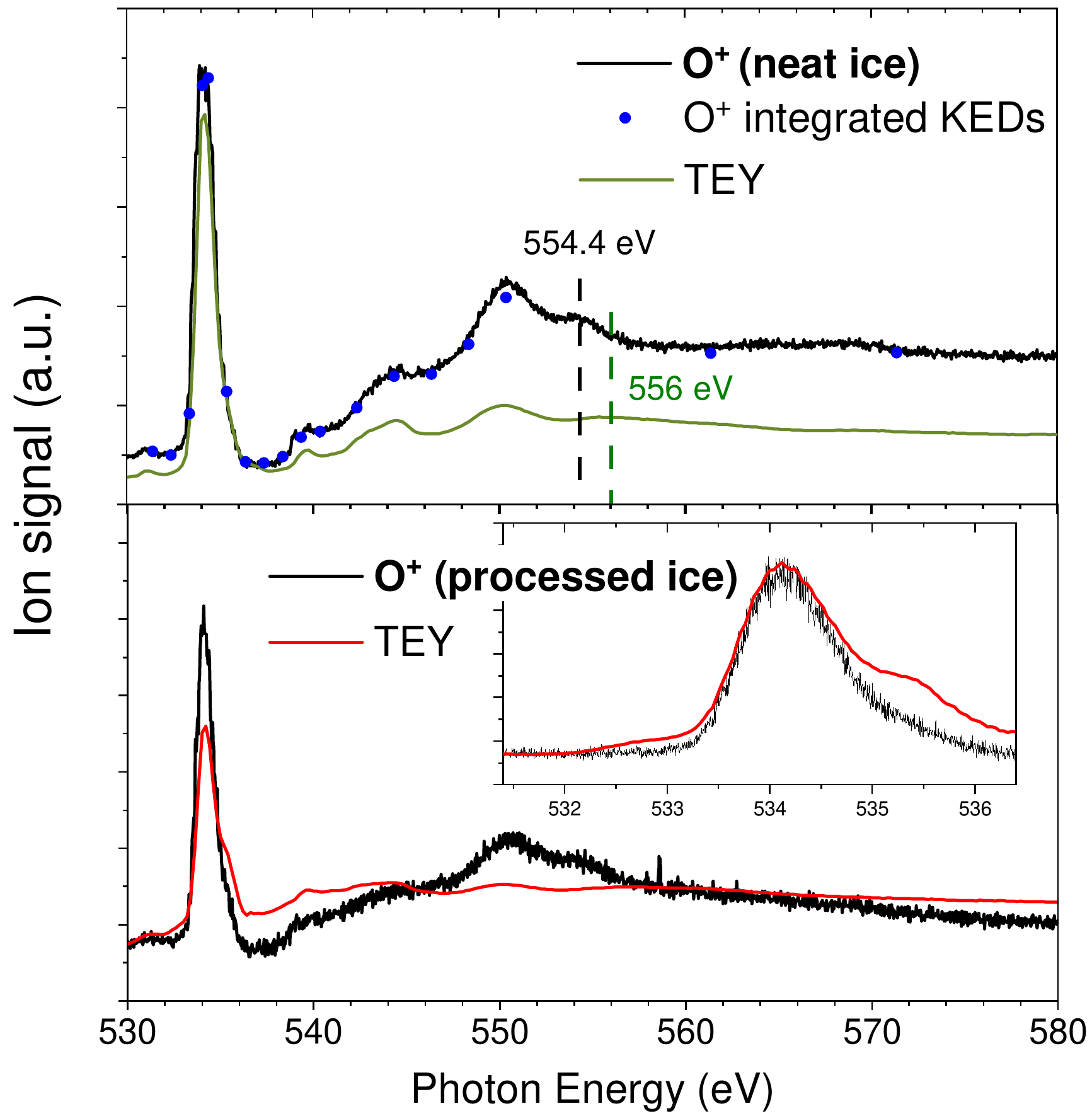}
    \caption{Photodesorption spectrum of O$^+$ for a neat ice (top panel) and a highly processed ice (bottom panel). Also shown on the top panel are the integrated KEDs for O$^+$ (blue dots) and the TEY (green line). The TEY is also shown in red in the bottom panel. The inset of the bottom panel is a zoom in the 1s$^{-1}\pi^*$ region, with the TEY scaled differently for comparison.} 
    \label{O+_psd}
\end{figure}

The photodesorption spectrum of O$^+$ is shown in fig. \ref{O+_psd}, for the neat CO ice in the top panel and the highly irradiated CO ice in the bottom panel. The neat ice photodesorption spectrum again resembles closely to the absorption. This time, however, the relative intensity of the 1s$^{-1}\pi^*$ to the near-IP region is lower than the TEY, and it is confirmed by the integrated KEDs. Production of O$^+$ is therefore higher above 538 eV than it is at 534.4 eV on the resonance. In addition, there is a feature occurring at 554.4 eV instead of 556 eV on the TEY. The highly irradiated ice desorption spectrum has a behaviour similar to the one observed for C$^+$: the O$^+$ spectrum remains more similar to the neat ice spectrum, although it is slightly more modified than the C$^+$ highly irradiated spectrum. A similar conclusion can be drawn: O$^+$ desorption through core-excitation of CO is more efficient than through core-excitation of other species. 

\begin{figure}
	\centering
    \includegraphics[trim={0cm 0cm 0cm 0cm},clip,width=\linewidth]{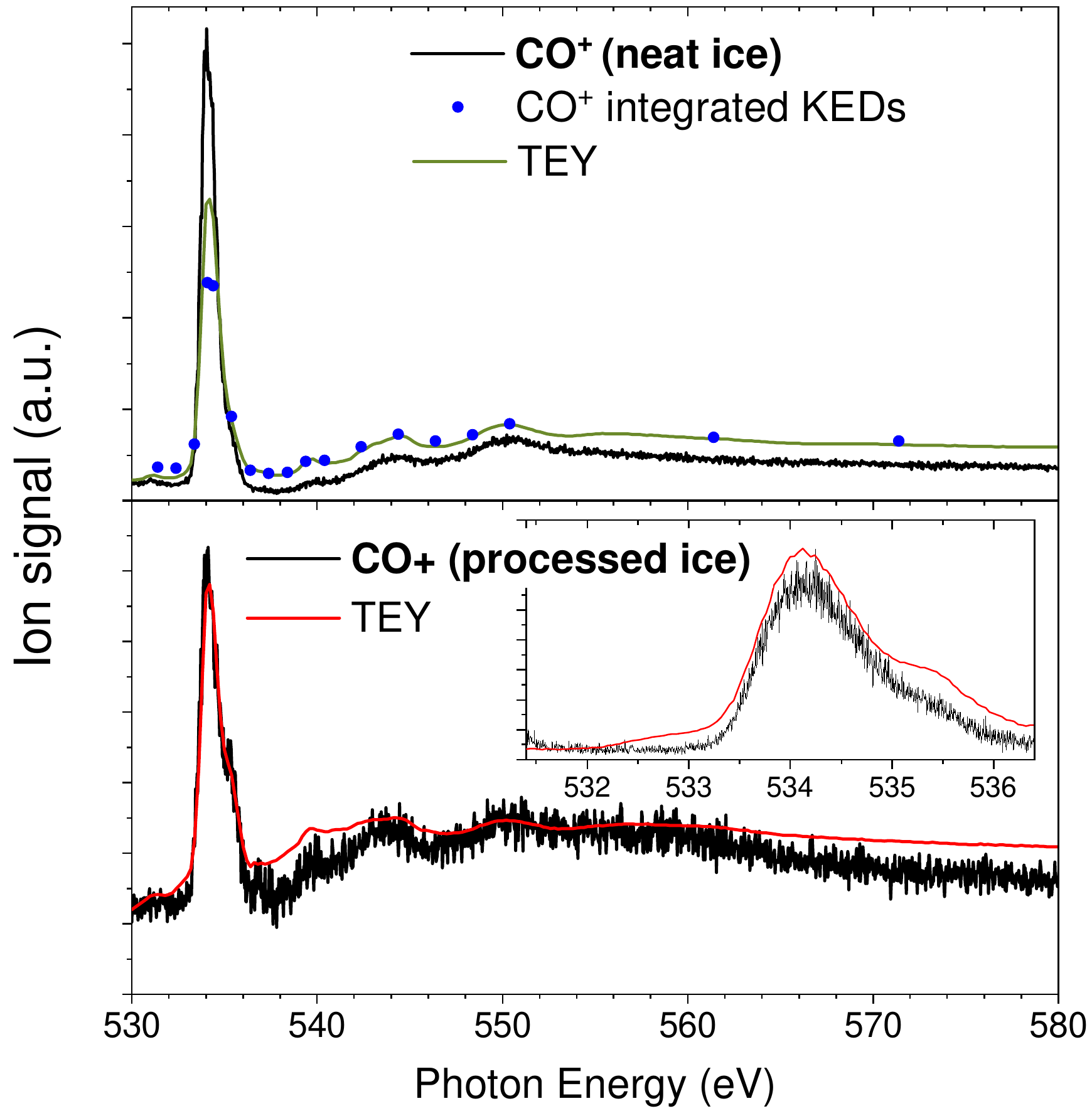}
    \caption{Photodesorption spectrum of CO$^+$ for a neat ice (top panel) and a highly processed ice (bottom panel). Also shown on the top panel are the integrated KEDs for CO$^+$ (blue dots) and the TEY (green line). The TEY is also shown in red in the bottom panel. The inset of the bottom panel is a zoom in the 1s$^{-1}\pi^*$ region.} 
    \label{CO+_psd}
\end{figure}

The photodesorption spectrum of CO$^+$, taken on the same ice as well, is shown in fig. \ref{CO+_psd}, again with the neat ice case for the top panel and the highly irradiated ice case for the bottom panel. For the neat ice spectrum displayed the 1s$^{-1}\pi^*$ peak is much more intense than the near-IP and continuum region compared with the TEY, but this is not confirmed by integrated KEDs, which indicate instead a slightly lower relative intensity. We must therefore attribute this observation to an artifact of charging and/or ageing. In the highly irradiated case (bottom panel of fig. \ref{CO+_psd}), the photodesorption spectrum also matches the absorption well, with discrepancies in the continuum slope that can be attributed to ageing. In the inset, we see that contrary to C$^+$ and O$^+$, the CO$_2$ absorption feature also appears as a weak shoulder (less pronounced than in the absorption spectrum) at 535.2 eV in the photodesorption spectrum. CO$^+$ seems to be an intermediate case between the behaviour of C$^+$ and O$^+$ on one hand, and a behaviour where the desorption spectrum would perfectly match the absorption on the other hand. 

\begin{figure}
	\centering
    \includegraphics[trim={0cm 0cm 0cm 0cm},clip,width=\linewidth]{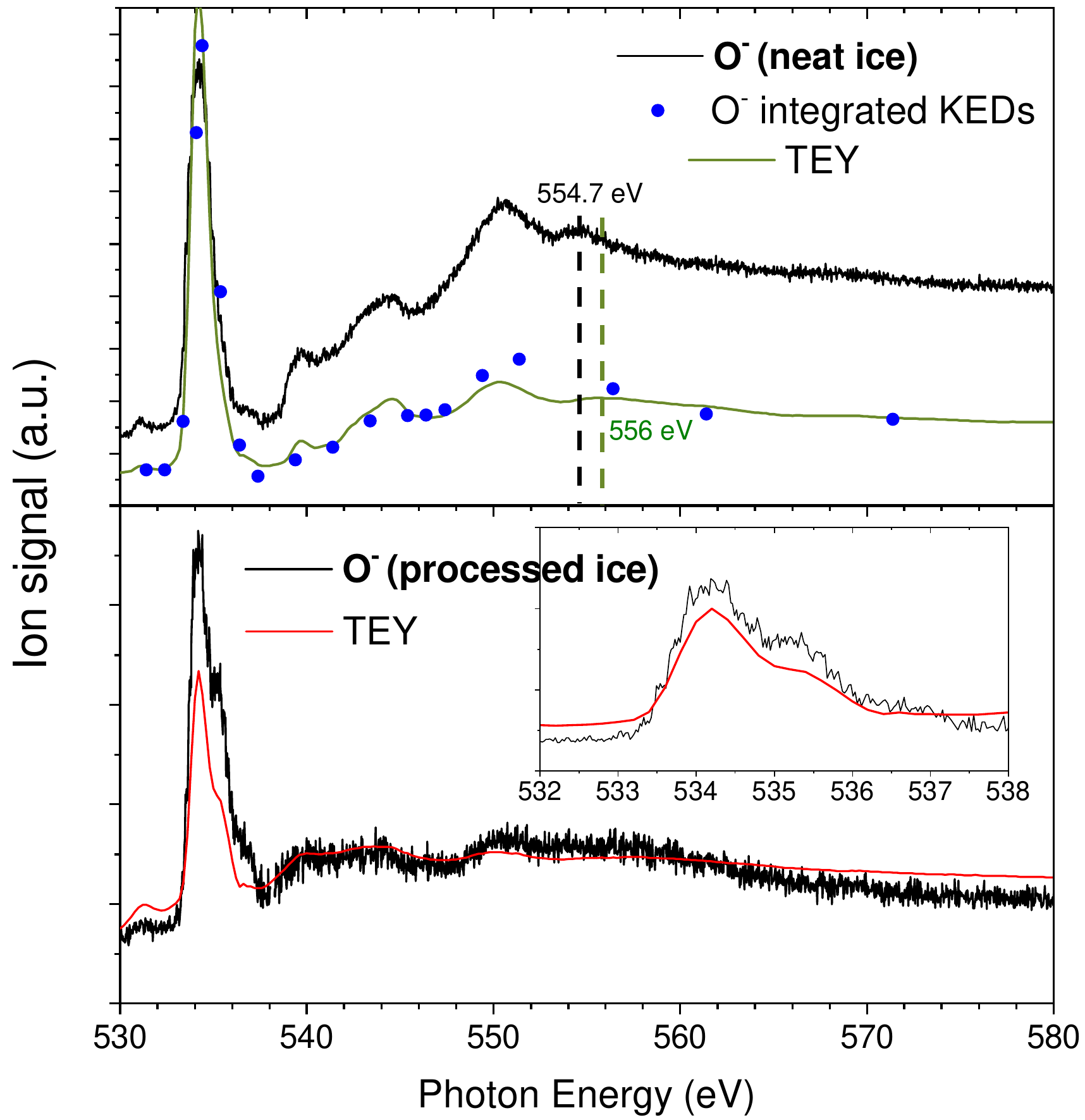}
    \caption{Photodesorption spectrum of O$^-$ for a neat ice (top panel) and a highly processed ice (bottom panel). Also shown on the top panel are the integrated KEDs for O$^-$ (blue dots) and the TEY (green line). The TEY is also shown in red in the bottom panel. The inset of the bottom panel is a zoom in the 1s$^{-1}\pi^*$ region.} 
    \label{O-_psd}
\end{figure}

We also looked at the desorption of anions. Only one anion was observed with significant intensity, O$^-$. C$^-$ was also observed with more than 30 times less signal, and the ion count was not sufficient to study it in more detail. The photodesorption spectrum of O$^-$ is shown on fig. \ref{O-_psd}, for a neat ice on the top panel (along with the integrated KEDs) and for a highly processed ice on the bottom panel. Looking at the photodesorption spectrum for the neat ice, there is no significant difference with the TEY, and this is confirmed by the integrated KEDs. The one difference with the TEY is the presence of a peak at 554.7 eV, very close to what is observed for O$^+$. For the processed ice, O$^-$ still follows the TEY quite well, with the distorted features in the near-IP region being clearly observed. This is in contrast with the behaviour of O$^+$ and C$^+$. Furthermore, if we zoom onto the 1s$^{-1}\pi^*$ (inset of the bottom panel), the CO$_2$ peak (shoulder at 535.2 eV) is quite strong, as is the following peak at 536.6 eV. Even for CO$^+$ the contribution of the non-CO peaks for the highly processed ice was not as strong. 

\begin{figure}
	\centering
    \includegraphics[trim={0cm 0cm 0cm 0cm},clip,width=0.8\linewidth]{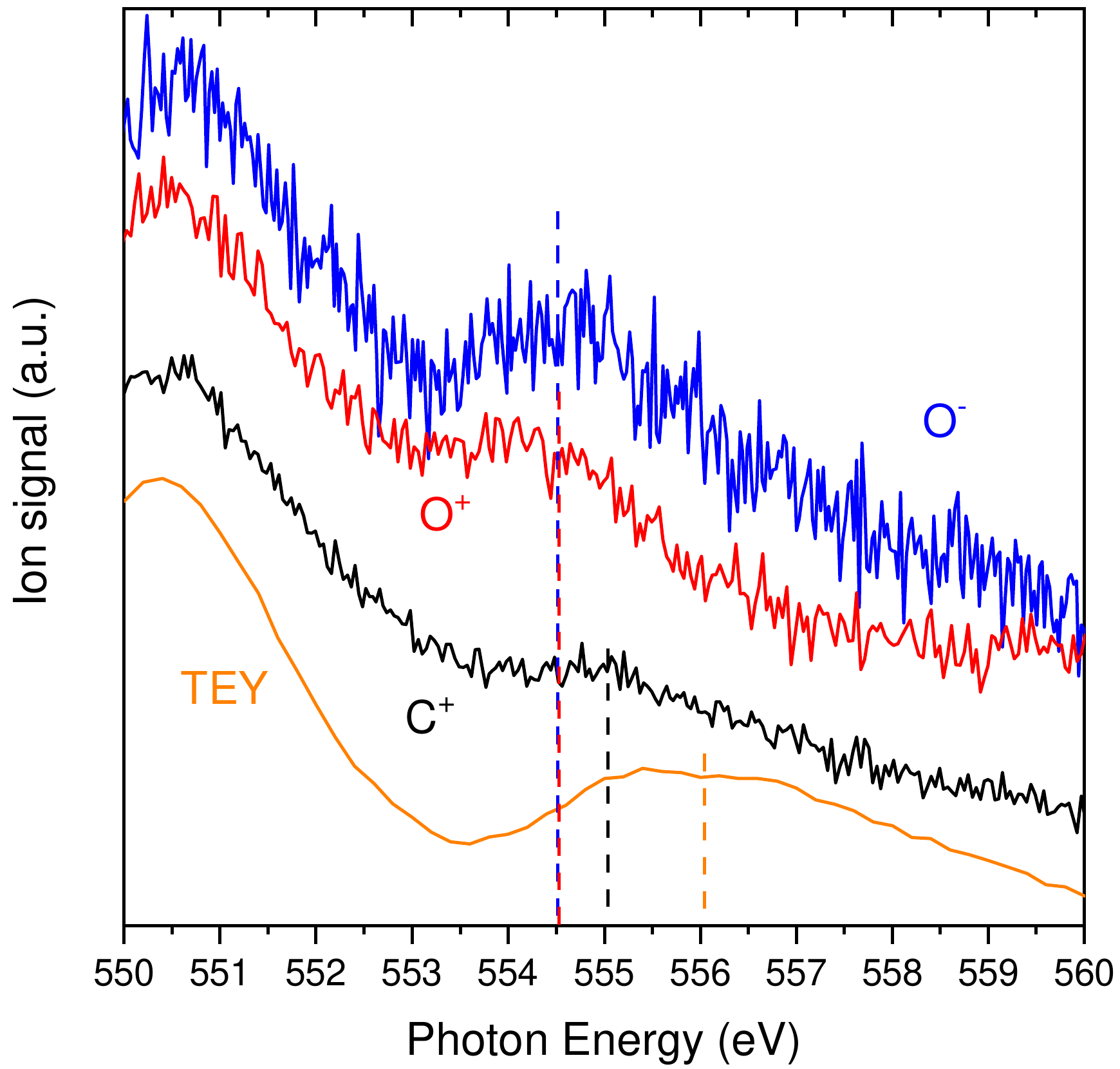}
    \caption{Zoom in the 550-560 eV region of the O$^+$, O$^-$ and C$^+$ photodesorption spectra and the TEY. The scaling is arbitrary and used to show clearly the peaks around 555 eV.} 
    \label{CO_X_double}
\end{figure}

Let us now dwell more on the small peak(s) that are observed around 555 eV in the different photodesorption spectra and the TEY. To have a better view of these peaks a zoom is made in fig. \ref{CO_X_double} of the 550-560 eV region for the TEY and O$^+$, O$^-$ and C$^+$ photodesorption spectra.  A peak appears in the absorption at 556 eV that we attributed to a 2h1e state following XPS data. One peak appears at 555 eV for C$^+$ but it is faint and there is some uncertainty on whether this red-shift could come from differences in the slope of the decreasing continuum. However the peaks are clearly visible and unambiguous in the spectra of O$^+$ and O$^-$, and red-shifted to around 554.5 eV. While the peak at 556 eV in the absorption may be attributed to a second shape resonance additional to the main one at 550 eV, this shape resonance overlaps with a series of doubly excited states (2h2e with a core excitation and a valence shake up excitation) occurring in the same region but with a lower total cross-section. In particular, the peaks observed in the desorption spectra can be attributed to a  1s$^{-1}$1$\pi^{-1}$2$\pi^{2}$ transition, predicted to occur in this region by theoretical calculations\cite{agren1984a} and observed experimentally by Stolte et al.\cite{stolte2001} in a study of O$^-$ emission from core-excited gas phase CO. Studies of ion yields like Stolte et al. and ours allow to reveal such double excitation transitions in the continuum because they have particularly high ion yield. In particular, anions have often been used as tracers of doubly-excited states above ionization thresholds, either in the valence or core region \cite{dujardin1989,dadouch1991,piancastelli2005}. Charge conservation requires that the dissociation of an initially positively charged state yielding an anion fragment also involves highly positively charged fragments, which makes these dissociation pathways minor (or even forbidden in some cases) in the case of core ionisation.  Doubly-excited states with low excitation cross-sections (therefore invisible in photoabsorption) but involving a neutral initial state thus become more visible in the ion yield spectra. Similarly, these 2h2e state could have a particularly high O$^+$ desorption yield which make them appear in the O$^+$ spectrum. 

\section{Discussion}

\subsection{Photodesorption mechanisms}

As in the case of water ice\cite{dupuy2018c,dupuy2020}, there is a first basic distinction we can make between two classes of photodesorption mechanisms: direct desorption (mediated by relaxation of the core-excited molecule itself) as opposed to X-ray induced Electron-Stimulated Desorption (XESD, mediated by scattering of the Auger electron in the ice and the following secondary events). Photodesorption spectra help in getting first insights into the direct desorption/XESD distinction: XESD-related mechanisms must necessarily have a photon energy dependence that follows the amount of electrons produced, i.e. the TEY. This is not necessarily the case for direct mechanisms, thus deviations of the photodesorption spectra from the TEY, provided they are clearly not measurement artifacts, indicate that such direct mechanisms are at play. We will discuss for the different desorbing species the direct/XESD distinction, but also suggest desorption mechanisms and compare with desorption induced by other types of irradiation.

\subsubsection{Neutral molecules.}

The photodesorption spectrum of neutral CO follows the TEY within the experimental noise and uncertainties related to ice ageing. This was also observed in Scheuerer et al.\cite{scheuerer1990}. It is also the case for CO$_2$, although here the noise of the signal is particularly high. These observations open the possibility that XESD is the dominant process for neutral desorption, as in the case of water ice\cite{dupuy2018c}. To comfort this hypothesis, it is useful to compare X-ray induced desorption and electron-stimulated desorption (ESD). 

For this we first need to derive the X-ray photodesorption yield per absorbed photon, through the procedure detailed in ref. \cite{dupuy2018c}. We can consider an absorption cross-section at 600 eV of $6\times 10^{-19}$ cm$^2$, which is the value for gas-phase CO \cite{berkowitz2002}. Measurements for solid CO do not exist, but far above threshold the cross-section can reasonably be considered to be only atomic in nature (neglecting EXAFS-type effects). The characteristic depth involved in photodesorption can be taken as $\sim$ 30 ML, which is the approximate mean range of the Auger electron in H$_2$O\cite{dupuy2018c} and should be reasonably similar in CO. The yield per absorbed photon obtained for the O 1s region, calculated at 600 eV, is 8.4 molecules per absorbed photon. This value is very similar to the yield for electron-induced desorption at 500 eV - about the energy of the Auger electron in the X-ray case - from CO ice\cite{dupuy2020b}. 

We also note that ESD of neutral CO from CO ice has a low energy threshold - around 6 eV\cite{rakhovskaia1995} (corresponding to the lowest lying excited state a$^3\Pi$). This is accessible to typical secondary electrons produced by scattering of the Auger electron, which have a kinetic energy peaking around 10 eV. Most of the energy initially carried by the X-ray photon goes into the Auger electron, and subsequently in the secondary events. Therefore, if desorption induced by these secondary events is energetically possible, there is a high chance that they will dominate the overall desorption process. All of these arguments lead us to the same conclusion as for water ice, which is that desorption of neutral species is dominated by XESD processes.

We can also compare X-ray photodesorption with VUV photodesorption of CO. For the $A^1\Pi-X^1\Sigma^+$ transition of CO in the VUV region, the photodesorption yield per absorbed photon is about 1, which is about an order of magnitude less than the X-ray yield. If we compare instead the photodesorption yield per deposited eV in the ice, we obtain 0.015 desorbed molecules per deposited eV in the X-ray range, and 0.1 desorbed molecules per deposited eV in the UV range. This indicates that VUV photodesorption - or more precisely desorption through the $A^1\Pi-X^1\Sigma^+$ electronic transition - is intrinsically more efficient (i.e. the photon energy is more efficiently converted into molecule desorption) than X-ray photodesorption. This can be understood considering that the $A^1\Pi-X^1\Sigma^+$ transition is non-dissociative. Excitation to the $A^1\Pi$ state will on average lead to the desorption of 1 CO molecule. On the other hand, scattering of the Auger electron can excite many states including ionic states and dissociative states. Energy will therefore be lost in bond breaking and ionisation, competing with the desorption channel. The present results suggest that states above the $A^1\Pi-X^1\Sigma^+$ state have desorption efficiencies (average number of desorbed molecules per excitation) lower than 1. 

\subsubsection{Fragment ions.}

Here we will discuss the desorption mechanisms of C$^+$, O$^+$, CO$^+$ and O$^-$. The photodesorption spectrum of these four ions from a neat CO ice (fig. \ref{C+_psd}-\ref{O-_psd}) follows the absorption relatively well, with some differences pointed out in the results section. This, however, does not mean that XESD necessarily dominates the desorption. Indeed, the spectra of C$^+$ and O$^+$ from highly irradiated CO ice are not heavily modified compared with the neat ice, whereas the absorption is. We can therefore conclude that XESD is not dominant for these two species. Furthermore, we also pointed out that desorption of these ions through core-excitation of CO is significantly more efficient than through core-excitation of CO$_2$ or other species contributing to the spectrum. The case of CO$^+$ is less clear, as the photodesorption spectrum resembles more, but not completely, to the absorption. For O$^-$ no significant difference is observed between desorption and absorption for either ices (except for the small peak around 554 eV discussed before). In these two cases it is not possible for now to conclude on the predominance of XESD or direct desorption. 

Looking at the desorption of ions in the XUV/deep valence region (25-60 eV) of condensed CO, as studied by Philippe et al.\cite{philippe1997}, is instructive. What has been observed is desorption of C$^+$ and O$^+$ with a photon energy threshold for desorption at 30 eV and a maximum yield around 50 eV, attributed to satellite states (2h1e states). These states are typically the final states reached after spectator Auger decay of core-excited CO (i.e. below threshold). The 5:1 ratio in intensity between C$^+$ and O$^+$ observed is close to the ratio we see here at 534.4 eV (3.5:1). This ratio is observed to decrease at higher energies (60 eV) in Philippe et al., although the study was not pushed to even higher energies where the contribution of double ionization (2h states) really dominates. 2h states are those reached after normal Auger decay, above threshold, and in that case the ratio C$^+$/O$^+$ that we observe is close to 2:1. For CO$^+$,  in the XUV study there is similarly desorption associated with satellite states excitation, with a threshold around 28 eV and increasing yield up to 50-60 eV. However, desorption of CO$^+$ is also observed at lower energies (14-25 eV) through excitation of resonances attributed to doubly excited Rydberg states (2h2e states). 

Therefore the desorption of C$^+$ and O$^+$ are fairly straightforward to explain through a direct desorption mechanism: below threshold, Auger decay leads to satellite/2h1e states which have been shown to trigger desorption of these ions in the XUV region, through the dissociative character of some of these highly excited states (most states other than X, A and B of CO$^+$ are dissociative in the gas phase). Above threshold, the 2h final states cause dissociation of the molecule through Coulomb explosion, yielding C$^+$ and O$^+$ fragments with a lot of kinetic energy, which are able to desorb. The Auger electrons released in the ice can similarly excite these same states, but they are rather high in energy, since a threshold at 30 eV has been observed in photodesorption. This is clearly not reachable by secondary electrons (0-20 eV typical kinetic energy). Even their excitation cross-section by the Auger electron must not be very high, even though such data is difficult to obtain. From these considerations, and considering additionally the deviations of the desorption spectra of C$^+$ and O$^+$ from the absorption TEY spectrum (fig. \ref{C+_psd} and \ref{O+_psd}) we can conclude that a direct desorption process, rather than XESD, that involves dissociation of highly excited CO$^{+*}$ or CO$^{++}$ is at the origin of C$^+$ and O$^+$ desorption.

The case of CO$^+$ is different. The mechanisms for CO$^+$ desorption in the first place are more complicated. The CO$^+$ ion needs to be formed with sufficient kinetic energy to overcome the desorption barrier: CO$^+$ desorption is not observed after single ionization to the ground state or a low excited state of CO$^+$\cite{philippe1997}. It was suggested, for excitation of doubly excited states, that a reduction of the intermolecular distance by forming a CO$^*$ + CO dimer followed by autoionization and repulsion could explain the desorption. For the satellite states it was proposed instead that the excited (CO$^+$)$^*$ could deexcite by ionizing a neighboring molecule, causing CO$^+$ - CO$^+$ repulsion\cite{philippe1997}. We could therefore imagine similar mechanisms for direct desorption in our case: formation of CO$^+$ - CO$^+$ pairs, either below or above threshold. Regarding the XESD vs. direct desorption question, the fact that the desorption threshold of CO$^+$ (14 eV) is at much lower energies than C$^+$ and O$^+$, thanks to the doubly-excited states, increases the probability that XESD plays a role. Indeed not just the Auger electron but also the high energy tail of the secondary electrons (15-20 eV) could conceivably excite these states as well. The available evidence from our experimental results and the literature does not exclude a significant XESD contribution to CO$^+$ desorption.  

The results obtained here on condensed CO are quite different from those obtained in the Menzel group on CO chemisorbed on metal surfaces \cite{treichler1991, weimar2000, feulner2000}. In these works, the desorption spectra of C$^+$ and O$^+$ (and O$^{2+}$) are very different from the absorption, and multielectron features play a much more prominent role. There are clear high energy features (around 550 and 570 eV) associated with 2h2e and 3h3e excitations that dominate the spectra. This is explained by the fact that more "simple" excitations are efficiently quenched by the substrate. Dissociation and desorption of ionic fragments thus require highly localized states like the doubly or triply excited states. Their result for CO$^+$, on the other hand, follow closely their absorption spectrum.   

In the case of O$^-$ desorption, we have even stronger incentives to consider XESD as a potentially dominant process. O$^-$ desorption can be induced by low-energy electrons with a threshold around 11 eV\cite{sanche1984}, through dissociative electron attachment (DEA). Ion-pair dissociation can also occur at higher electron energies. Secondary electrons can therefore readily participate to O$^-$ desorption. Anion desorption by DEA from condensed CO also mostly yields O$^-$ desorption and ten times less C$^-$, with also a higher electron energy threshold for C$^-$, which is in agreement with our observations. However, direct desorption processes cannot be entirely ruled out.

It may seem surprising to suggest that direct processes, which involve the dissociation of CO$^+$ and CO$^{++}$, would yield anions with an efficiency that can compete with DEA and ion-pair dissociation from secondary and Auger electrons. In gas phase core ionization, O$^-$ is observed below but also above the core ionization threshold\cite{stolte2001,hansen2002}. This requires the formation of the ion pair C$^{3+}$ + O$^-$, a pathway that should have a low probability but is viable. Such dissociation pathways are usually three to four orders of magnitude lower than dissociation to only cation and neutral fragments. Despite this, we observed in the case of water ice\cite{dupuy2020} desorption of H$^-$ and O$^-$ with yields only an order of magnitude lower than their cation counterpart, along with clear indications that their desorption is not dominated by an XESD process but by a direct process. Indeed, dissociation of singly or doubly ionized molecules to form an anion is easier in condensed phase where processes like intermolecular coulombic decay (ICD) or charge transfer can split positive charges over several molecules. While in the present case there is no indication of a dominant direct process in the desorption of O$^-$ anions, the fact that in the case of anion desorption from water ice direct processes are dominant suggest they could be dominant here as well. This prevents us from definitively concluding that O$^-$ desorption from CO ice is dominated by XESD. 

\subsubsection{Large ions.}

Large ions and ion clusters were observed to desorb from the ice (table \ref{mspec_CO_32+}). Desorption of these larger ions is hard to explain by simple ionization of their neutral equivalent formed in the ice. The reason for that is similar to the arguments explaining why simple ionization of CO does not lead to CO$^+$ desorption. Desorption of these ions could conceivably occur from fragmentation of bigger molecules, but then the lighter fragments would take away most of the dissociation energy, and break up into two large fragments is not very likely. Instead, the most likely explanation is that these ions desorb immediately upon their formation via ion-neutral reactions, owing to the exothermicity of the reaction. This hypothesis was suggested by Philippe et al.\cite{philippe1997} to explain the desorption of ions slightly larger than CO$^{+}$ but can be extended to the larger cations. For example C$_2$O$^+$ can be formed via C$^+$ + CO or (CO$^+$)$^*$ + CO (exothermicity 2.8 eV in the gas phase\cite{bowers1975}), and C$_3$O$_2^+$ via C$_2$O$^+$ + CO (exothermicity more than 6 eV in the gas phase\cite{philippe1997}). Exothermic addition reactions of C$^+$, CO$^+$ or C$_2$O$^+$ to existing C$_x$O$_y$ species could explain the variety of ions observed in desorption. This would also explain why the desorption yield of CO$_2^+$ is so low: simple ionization of the abundant CO$_2$ in the ice does not lead to desorption and CO$_2^+$ is not created via ion-neutral reactions (at least not with significant excess energy). 

\subsection{Photochemistry}

Photodesorption and photochemistry are intertwined, as the above discussion of desorption mechanisms should show. Although the primary focus of this work is on desorption, it is interesting to discuss photochemistry of CO ice as revealed by the techniques employed here, namely X-ray absorption spectroscopy (XAS) and Photon-stimulated ion desorption. These techniques are much less employed to study molecular ice chemistry than the usual techniques of IR spectroscopy and TPD but provide complementary information. X-ray absorption spectroscopy allows to study the presence of species that are undetected in IR spectroscopy, most notably O and O$_2$ here. Photon-stimulated ion desorption reveals a very large number of desorbing cations, much larger than the number of neutral species detected in the solid phase by IR spectroscopy and TPD\cite{ciaravella2016}. 

CO photochemistry can be initiated by mainly three types of channels: C and O radicals resulting from CO dissociation (directly, through dissociative electron attachment or through electron-ion recombination, depending on the energy range considered), reaction of electronically excited CO$^*$, and ion reactions involving CO$^+$ ions and fragments like C$^+$ and O$^+$ (above the respective ionization and dissociative ionization thresholds). 

The most relevant work to compare with here is the one of Ciaravella et al. \cite{ciaravella2012,ciaravella2016}, since it is the only one having studied directly X-ray photochemistry. Using transmission IR spectroscopy and TPD, they found production of several C$_x$ (x = 3, 5, 8, 9), C$_x$O (x = 2 - 6) and C$_x$O$_2$ (x = 1, 3, 5, 7) molecules, similar to those we saw in the form of cations. The most abundant detected molecules are CO$_2$, C$_3$O$_2$, C$_3$ and C$_2$O, coherent with the results we obtain for cations and neutral desorption. Most studies of the photochemistry/irradiation chemistry of CO have been done using hydrogen lamps (VUV photons) \cite{gerakines1996,loeffler2005}, energetic electrons \cite{jamieson2006} or energetic ions \cite{jonkman1981,haring1984,chrisey1990, trottier2004,loeffler2005,palumbo2008,seperueloduarte2010}. Since the processes leading to chemistry are mostly similar, these studies can be compared to soft X-ray photochemistry as well. The studies made using VUV photons ($\leq$11 eV) mentioned before \cite{gerakines1996,loeffler2005} are interesting because the only starting point of chemistry is the reaction of CO$^*$ with CO to form CO$_2$ and C. Since no O atoms are produced in this reaction, this somewhat restricts the chemical possibilities, but the major products of CO chemistry are suboxides like C$_2$O or C$_3$O$_2$, which are already observed with VUV irradiation and well explained by reactions of the type C + CO $\longrightarrow$ C$_2$O and C$_2$O + CO $\longrightarrow$ C$_3$O$_2$. Species observed by IR spectroscopy in these studies are C$_2$O, CO$_2$, C$_3$O$_2$ (in both studies), C$_3$, C$_3$O (ref. \cite{gerakines1996}), C$_5$O$_2$ and C$_7$O$_2$ (ref. \cite{loeffler2005}) and can all be explained by the above reactions. The study of Jamieson et al. \cite{jamieson2006}, made using 5 keV electron bombardment of CO ice, has a very thorough discussion of CO radiation-induced chemistry. They detected C$_x$ (x = 3, 6), C$_x$O (x = 2-7) and C$_x$O$_2$ (x = 1, 3-5). These authors also proposed a chemical network and attempted to derive rate constants for each reactions based on their results. However, they made the assumption that CO chemistry was essentially based on the CO$^*$ initial reaction, arguing that they saw no trace of O atom chemistry and that ions were negligible. In our study we have through XAS the evidence that there is a considerable amount of O and O$_2$ accumulated in the ice (which cannot be observed through IR spectroscopy). While it is not the goal of this study to attempt a precise quantification of the XAS spectra obtained, from the peak areas in fig. \ref{CO_TEYs} for scan 4, one can estimate that O$_2$ amounts to $\sim$ 3\% of the total CO + CO$_2$ of the ice, while O radicals amount to $\sim$ 0.4\%.  

The results discussed above on ion desorption also emphasize that ion-neutral reactions do take place, although here any quantitative assessment is not possible. The species detected in our work in the form of desorbed cations include all the species already detected in the solid phase in neutral forms in other works, but also many more. This is allowed by the very high sensitivity of ion detection in the gas phase. However ion desorption remains an indirect method, and thus obtaining meaningful information out of the relative intensities observed is very much non-straightforward. As an example, desorption of CO$_2^+$ is quite low while CO$_2$ is the most abundant photoproduct in the ice. A detailed understanding of desorption mechanisms would therefore be required for a more in-depth analysis of the relative intensities of desorbing cations. Despite these difficulties, the very high sensitivity of the technique - to the point where trace pollutants in the ice can be detected, as discussed in section 3.3.1 - makes it interesting from a qualitative point of view.

\subsection{Astrophysical implications}

Let us now briefly discuss the application of our results in the context of astrochemistry. We first conclude, as we did in the case of water ice X-ray desorption\cite{dupuy2018c}, that neutral desorption is much higher than ion desorption. The estimated desorption yield of C$^+$ - the most abundant ion - at 550 eV is 2 $\times 10^{-5}$ ion/incident photon, four orders of magnitude lower than neutral CO. The desorption of neutral CO$_2$ is also two orders of magnitude lower than neutral CO, and we therefore focus on the desorption of neutral CO which is quantitatively the highest. We also note that the observed desorption of very large cations (although with very low desorption yields) cannot be simply extrapolated to an astrophysical context considering the very high photon doses used here. This may also affect the desorption of neutral CO to some degree.

We can, as we did previously for water ice X-ray desorption\cite{dupuy2018c}, derive desorption yields per "average" photon for different X-ray environments relevant to astrophysics. The procedure is described in detail in our previous work\cite{dupuy2018c}.

It is assumed first that the photodesorption yield will follow the absorption of solid CO from 600 eV to 10 keV. This assumption is probably slightly wrong, because after a few hundreds of eV the photoelectron plays a role comparable to the Auger electron (around 800 eV the photoelectron has an energy of $\sim$ 250 eV, and the electron-stimulated desorption yield is maximal at this energy). The second hypothesis is that the absorption of solid CO in this energy range is the same as the absorption of gas phase CO. As argued previously, this is reasonable because the absorption of the core shells is mostly atomic in nature in this range, and hardly perturbed by external factors. An extrapolated photodesorption yield of CO up to 10 keV is thus derived. In this study, we have restricted our investigation to the O 1s edge, neglecting the C 1s edge. From the astrophysics context point of view this is justified because low photon energies are highly attenuated in the shielded regions where ices are found, and thus do not contribute significantly to the average desorption yield. Photon energies below 520 eV were thus neglected in the calculation of the extrapolated yields.

This extrapolated yield is weighted-averaged with X-ray spectra presented in our previous work\cite{dupuy2018c}. We used measured X-ray spectra of two different astrophysical sources, one being the young star TW Hya (representative of the type of spectrum illuminating a protoplanetary disk) and the other the MK-231 (representative of an average spectrum that could illuminate a molecular cloud close by). We calculated the attenuation caused by dust and gas to derive local X-ray spectra relevant to locations deep into the clouds, where icy grains are located. The derived average photodesorption yields are given in table \ref{CO_astro_yields}. Extra caution should be taken when considering the validity of these numbers, because as we saw the yield of CO depends on the ageing of the ice.

\begin{table}
	\centering
	\caption[Average photodesorption yields of intact CO for the X-ray spectra of different regions at different attenuations]{Average photodesorption yields of intact CO (CO molecules per incident photon of "average" energy for the given environment) for the X-ray spectra of different regions at different attenuations.}
	\def\arraystretch{1.2}
		\begin{tabular}{c c c}
				   				  & TW Hya (young star) & MKN231 (ULIRG) \\
		\hline
		Source spectrum           & $8.8 \times 10^{-2}$ & $2.9 \times 10^{-2}$	\\
	    n$_H$ = 10$^{21}$ cm$^2$  & $7.8 \times 10^{-2}$ & $2.3 \times 10^{-2}$ \\
	    n$_H$ = 10$^{22}$ cm$^2$  & $3.9 \times 10^{-2}$ & $4.6 \times 10^{-3}$	\\
	    n$_H$ = 10$^{23}$ cm$^2$  & $3   \times 10^{-3}$ & $5   \times 10^{-4}$	\\
	    n$_H$ = 10$^{24}$ cm$^2$  & $3.2 \times 10^{-4}$ & $1.7 \times 10^{-4}$	\\
		\end{tabular}	
		\label{CO_astro_yields}
\end{table} 

The evolution of these yields with the density traversed and with the spectrum are mostly that same as for water ice\cite{dupuy2018c} as progressively harder X-rays contribute dominantly to the spectrum, the average photodesorption yield becomes lower. One important difference with water is that CO freezes out on grains at much lower temperatures ($<$ 20 K), and thus deeper into clouds. Looking at the numbers from the table, it is likely that the average X-ray photodesorption yield for CO at a relevant location in a disk will be lower (in the few 10$^{-3}$ molec ph$^{-1}$ range) than the average yield for UV photons (10$^{-2}$ molec ph$^{-1}$). In such a location UV from the disk will probably not penetrate, but X-ray photodesorption will compete with cosmic ray desorption and photodesorption by cosmic ray induced secondary UV photons. More precise modelling would be necessary to determine the outcome of this competition. If the inside of the disk is shielded against cosmic rays, as was suggested recently\cite{cleeves2015}, then the competition to X-rays would be much lower.

\begin{table*}
	\centering
	\caption{Summary of the discussion on X-ray photodesorption mechanisms from CO ice. The order of magnitude of the photodesorption yields are also indicated.}
	\def\arraystretch{1.2}
		\begin{tabular}{>{\centering\arraybackslash}m{2.5cm} ||
		>{\centering\arraybackslash}m{3cm} |
		>{\centering\arraybackslash}m{2.5cm} |
		>{\centering\arraybackslash}m{2.5cm} |
		>{\centering\arraybackslash}m{2.5cm} |
		>{\centering\arraybackslash}m{2.5cm} |
		}
				   				  & Neutrals & C$^+$/O$^+$ & CO$^+$ & O$^-$ & Large cations \\
		\hline \hline
		Direct process & & Direct dissociation & \multirow[c]{3}[8]{=}{\centering Excited dimer and/or CO$^+$ ion pair repulsion} & Direct dissociation & \multirow[c]{3}[8]{=}{\centering Ion-neutral exothermic reactions} \\
		\cline{1-3}\cline{5-5}
		Auger scattering (XESD) & \multirow[c]{2}{=}{\centering Low-energy excitations (exact mechanism unclear)} & (Direct dissociation) & & (Direct dissociation) & \\
		\cline{1-1}\cline{3-3}\cline{5-5}
		Secondary electrons (XESD) & & & & DEA & \\
		\hline\hline
		Yield (molec ph$^{-1}$) & 10$^{-3}$ - 1 & 10$^{-5}$ & 10$^{-6}$ & 10$^{-6}$ & 10$^{-9}$ - 10$^{-7}$ \\
		\end{tabular}	
		\label{mechanisms}
\end{table*} 

\section{Conclusion}

We presented a detailed investigation of desorption induced by X-rays in CO ice. Most notably we observed desorption of neutral CO and CO$_2$, with neutral CO being by far the most abundant desorption product. The absolute desorption yields of neutral CO have therefore been extrapolated to astrophysically relevant yields.

We also observed desorption of a large variety of cations. The cation mass spectrum for low masses, corresponding mostly to fragments, gives insights into the minor, "parasitic" components of the ice: isotopes (here $^{12}$C and $^{18}$O instead of $^{13}$C and $^{16}$O), presence of hydrogen and possibly nitrogen (in very small amounts) impurities. For larger masses the spectrum is a testimony of the rich chemistry induced by X-rays in the ice, with series of C$_x^+$, C$_x$O$^+$, C$_x$O$_2^+$... cations being observed up to the detection limit. The evolution of the X-ray absorption spectrum is also a probe of chemistry, with both pre-edge peaks that can be attributed to the formation of new species and modifications of the post-edge region. These two probes provide information complementary to more common probes of ice chemistry. 

We also discussed the desorption mechanisms of the different species observed. This discussion is summarized in table \ref{mechanisms}. First, we attempt to distinguish between direct desorption and X-ray induced Electron Stimulated Desorption (XESD). Considerations related to the differences between absorption and desorption spectra, the energy threshold for desorption of the different species, and quantitative comparisons to direct ESD measurements, allowed us to draw some conclusions. Although most of the initial energy of the X-ray photon goes into the Auger electron, and therefore into XESD processes, XESD does not dominate desorption for all species. This is clearly not the case for C$^+$ and O$^+$ desorption. On the other hand, neutral desorption is dominated by XESD. For other species (notably CO$^+$ and O$^-$, but this also includes large cations), we could not draw definitive conclusions. O$^-$ desorption can occur both through DEA of low energy secondary electrons and direct dissociation of highly excited CO$^+$/CO$^{++}$. Large cations presumably desorb upon formation through exothermic ion-neutral reactions. 

\section*{Author contributions}

All authors participated to the synchrotron experiments from which the data of this manuscript originates. R.D. analysed the data and wrote the manuscript. M.B., G.F. and J.-H.F. provided extensive input on data analysis and the manuscript. P.J., J.-H.F. and M.B. designed the experimental set-up. R.D., M.B., G.F., J.-H.F., T.P., L.P., X.M. and P.J. contributed to the construction and characterization of the experimental set-up. J.-H.F., X.M., M.B. and V.B. acquired the necessary funding and beamtime.  

\section*{Conflicts of interest}

There are no conflicts to declare.

\section*{Acknowledgements}

This work was done in collaboration and with financial support by the European Organization for Nuclear Research (CERN) under the collaboration agreement KE3324/TE. We acknowledge SOLEIL for provision of synchrotron radiation facilities under the projects 20161406 and 20181140 and we thank Nicolas Jaouen and the SEXTANTS team for their help on the beamline. This work was supported by the ANR PIXyES project, grant ANR-20-CE30-0018 of the French Agence Nationale de la Recherche. This work was also supported by the Programme National "Physique et Chimie du Milieu Interstellaire" (PCMI) of CNRS/INSU with INC/INP co-funded by CEA and CNES. Financial support from the LabEx MiChem, part of the French state funds managed by the ANR within the investissements d'avenir program under reference ANR-11-10EX-0004-02, and by the Ile-de-France region DIM ACAV program, is gratefully acknowledged.

%%%END OF MAIN TEXT%%%


\balance


%%%REFERENCES%%%
\providecommand*{\mcitethebibliography}{\thebibliography}
\csname @ifundefined\endcsname{endmcitethebibliography}
{\let\endmcitethebibliography\endthebibliography}{}

\end{document}